\documentclass[onecolumn,12pt,nofootinbib]{revtex4-1}
\usepackage{graphicx}

\newcommand\aj{{AJ}}%
\newcommand\araa{{ARA\&A}}%
\newcommand\apjl{{ApJ}}%
\newcommand\apjs{{ApJS}}%
\newcommand\aap{{A\&A}}%
\newcommand\bain{{Bull.~Astron.~Inst.~Netherlands}}%
\newcommand\qjras{{QJRAS}}%
\newcommand\mnras{{MNRAS}}%
\newcommand\ssr{{Space~Sci.~Rev.}}%
\newcommand\zap{{ZAp}}%
\newcommand\azh{{AZh}}%
\newcommand\jcap{{J. Cosmology Astropart. Phys.}}%
\newcommand\pasj{{PASJ}}%

\def\lsim{\mathrel{\raise.3ex\hbox{$<$\kern-.75em\lower1ex\hbox{$\sim$}}}}
\def\gsim{\mathrel{\raise.3ex\hbox{$>$\kern-.75em\lower1ex\hbox{$\sim$}}}}

\begin{document}

\hspace{11.5cm} \parbox{48cm}{FERMILAB-PUB-16-157-A}~\\

\bibliographystyle{apsrmp4-1}

\title{A History of Dark Matter}
%\title{A Short History of Dark Matter}
\author{Gianfranco Bertone$^1$ and Dan Hooper$^{2,3}$}

\affiliation{$^1$GRAPPA, University of Amsterdam, Netherlands}
\affiliation{$^2 $Center for Particle Astrophysics, Fermi National Accelerator Laboratory, USA}
\affiliation{$^3$Department of Astronomy and Astrophysics, The University of Chicago, USA}

\date{\today}

\begin{abstract}

Although dark matter is a central element of modern cosmology, the history of how it became accepted as part of the dominant paradigm is often ignored or condensed into a brief anecdotical account focused around the work of a few pioneering scientists. The aim of this review is to provide the reader with a broader historical perspective on the observational discoveries and the theoretical arguments that led the scientific community to adopt dark matter as an essential part of the standard cosmological model.

\end{abstract}

\maketitle

\tableofcontents

\cleardoublepage
%\pagenumbering{arabic}

\section{Preface}
\label{foreward}

Dark matter plays a central role in our understanding of modern cosmology. But despite its significance, many of the scientists active in this area of research know relatively little about its interesting history, and how it came to be accepted as the standard explanation for a wide variety of astrophysical observations. Most publications and presentations on this topic -- whether at a technical or a popular level -- either ignore the long history of this field or condense it into a brief anecdotal account, typically centered around the work on galaxy clusters by Fritz Zwicky in the 1930s and on galactic rotation curves by Vera Rubin in the 1970s. Only a small number of scientists, and an even smaller number of historians, have endeavoured to systematically analyze the development of the dark matter problem from an historical perspective, and it is surprisingly hard to find articles and books that do justice to the fascinating history of dark matter.

The aim of this article is to provide a review of the theoretical arguments and observations that led to the establishment of dark matter among the pillars of modern cosmology, as well as of the theories that have been proposed to explain its nature. Although we briefly discuss some early ideas and recent developments, the focus of this review is the 20th century, beginning with the first dynamical estimates of dark matter's abundance in the Universe, and to its role in the current standard cosmological model, and the strategies that have been pursued to reveal its particle nature.

The first part of this article is largely based on the analysis of primary sources, mainly scanned versions of scientific articles and books published in the 19th and 20th centuries, freely accessible via NASA ADS and the Internet Archive Project. We study the emergence of the concept of dark matter in the late 19th century and identify a series of articles and other sources that describe the first dynamical estimates for its abundance in the known Universe (Chapter~\ref{prehistorychapter}). We then discuss the pioneering work of Zwicky within the context of the scientific developments of the early 20th century. And although his work clearly stands out in terms of methodology and significance, we find that his use of the term ``dark matter'' was in continuity with the contemporary scientific literature. We then go on to follow the subsequent development of the virial discrepancy that he discovered, with particular emphasis on the debate that took place around this issue in the 1960s (Chapter~\ref{sec:clusters}).

The second part of this article focuses on more recent developments, which gave us the opportunity to complement the analysis of the primary sources with extensive discussions with some of the pioneering scientists who contributed to the advancement of this field of research. We discuss the history of galactic rotation curves, from the early work in the 1920s and 1930s to the establishment of flat rotation curves in the 1970s, placing the famous work of Bosma and Rubin and collaborators in 1978 within the broader context of the theories and observations that were available at that time (Chapter~\ref{Chap:rotcurves}). We then discuss the theories that have been put forward to explain the nature of dark matter, in terms of fundamental particles (Chapter~\ref{particles}), astrophysical objects (Chapter~\ref{baryonicdm}), or manifestations of non-Newtonian gravity or dynamics (Chapter~\ref{modgrav}). 

Finally, we discuss how the emergence of cosmology as a science in the 1960s and 1970s, the advent on numerical simulations in the 1980s, and the convergence between particle physics and cosmology, led most of the scientific community to accept the idea that dark matter was made of non-baryonic particles (Chapter~\ref{chap:cosmology}), and prompted the development of new ideas and techniques to search for dark matter candidates, many of which are still being pursued today (Chapter~\ref{sec:searches}). 
 
One of the main difficulties in reconstructing the history of dark matter is that the key developments took place in a continuously changing landscape of cosmology and particle physics, in which scientists were repeatedly forced to revise their theories and beliefs. The authors of this review are not professional historians, but scientists writing for other scientists. And although we have taken great care in reconstructing the contributions of individuals and groups of scientists, we have little doubt that that our work falls short of the standards of the historical profession. We nevertheless hope that this article will contribute to a better understanding and appreciation of the history of dark matter among our fellow astronomers and physicists, and that it will foster an interest among professional historians in this rich and fascinating field of research.

\section{Prehistory}
\label{prehistorychapter}

%Astronomers and philosophers of all times have speculated about the possible existence of invisible forms of matter in the Universe. Even before the modern concept of dark matter was introduced, the more fundamental question was asked of whether something else existed beside the planets and stars that could be observed at night with the naked eye. 
%
%The elegant but flawed aristotelian cosmology, which dominated much of the history of Western thought, before the advent of modern science, actively discouraged these speculations, making a strong distinction between the immutable heavens, on which planets are seen moving against the vault of fixed stars, and the Earth, center of the Universe. Attempts to modify this view have met strong resistance, as it is well known, but alternative ideas have been put forward by philosophers and scientists before, during and after the golden age of aristotelian cosmology. 
%
%In this chapter, we will we briefly present some of these alternative ideas, in order to set the stage for the discussion of the discovery of dark matter. The pragmatic reader who is interested in modern developments only, can skip the first two sections, but we encourage her or him to read the last section, where we review the pioneering work of Kapteyn, Jeans and Oort, and their attempts to infer the abundance of dark matter form dynamical observations of nearby stars.

\subsection{From Epicurus to Galileo}

Throughout history, natural philosophers have speculated about the nature of matter, and even have contemplated the possibility that there may be forms of matter that are imperceptible -- because they were either too far away, too dim, or intrinsically invisible. And although many of the earliest scientific inquiries were less than rigorous, and often inseparable from philosophy and theology, they reveal to us the longevity of our species' desire to understand the world and its contents.

Although many early civilizations imagined their own cosmological systems, it was arguably the ancient Greeks who were the first to attempt the construction of such a model based on reason and experience.  The atomists, most famously Leucippus and Democritus who lived in the 5th century BCE, were convinced that all matter was made of the same fundamental and indivisible building blocks, called atoms, and that these atoms were infinite in number, as was the infinite space that contained them. Epicurus (341 BCE -- 270 BCE) further suggested in his ``Letter to Herodotus'' that an infinite number of other worlds existed as well, ``some like this world, others unlike it''\footnote{\small Epicurus, Letter to Herodotus (c. 305 BCE), Extracted from Diogenes Laertius, Lives of Eminent Philosophers, trans. R. D. Hicks, vol. 2 (1925).}. Others speculated about unobservable matter that might be found within our own Universe. For example, the Pythagorean Philolaus conjectured the existence of the celestial body {\it Antichthon}, or counter-earth, which revolves on the opposite side of the ``central fire'' with respect to the Earth~\cite{kragh2006conceptions}.

The cosmological model of Aristotle -- which would dominate discourse throughout the Middle Ages -- provided an elegant construction, in which the location of the Earth was fixed to the center of an immutable Universe. This model offered what seemed to many to be strong arguments against the existence of invisible or unknown forms of matter. Even the striking appearance of comets, which obviously had no place in Aristotle's highly organized hierarchy of celestial spheres, was dismissed as an atmospheric phenomenon, a belief that continued to be held until Tycho Brahe measured the (absence of) parallax for a comet in 1577.

Although many offered challenges to the orthodoxy of Aristotelian cosmology, these attempts were not met without resistance. The statue of Giordano Bruno in Campo de' Fiori in downtown Rome serves as a reminder of the dangers that were inherent in such departures from the strict Aristotelian worldview embraced by the Catholic Church. It was at the location of that statue that Bruno was burned at the stake in 1600 by the Roman Inquisition, after being convicted on charges that included the holding of a heretical belief in the existence of infinite other worlds.

It was arguably Galileo -- who himself had his share of trouble with the inquisition -- who did the most to break the hold of Aristotelian cosmology. By pointing his telescope toward the sky, Galileo saw much that had been previously imperceptible. Among his many other discoveries, he learned that the faint glow of the Milky Way is produced by a myriad of individual stars, and that at least four satellites, invisible to the naked eye, are in orbit around Jupiter. Each of these observations encapsulate two lessons that remain relevant to dark matter today. First, the Universe may contain matter that cannot be perceived by ordinary means. And second, the introduction of new technology can reveal to us forms of matter that had previously been invisible.

%who settled to the eyes of scientifically-inclined minds the question of whether the Universe really contained more than meets the eye: by pointing his telescope to the sky, he realized that the faint glow of the Milky Way was produced by miryads of stars, and that at least 4 satellites, invisible to the naked eye, orbit around Jupiter. 

\subsection{Dark Stars, Dark Planets, Dark Clouds}

The course of science, and of astronomy in particular, was transformed in 1687 when Isaac Newton published his treatise {\it Philosophi\ae~Naturalis Principia Mathematica}. Newton's Laws of motion and Universal Gravitation provided scientists with new and formidable tools which, among many other things, enabled them to determine the gravitational mass of astronomical bodies by measuring their dynamical properties.  

In 1783, John Michell, also famous for inventing the torsion balance for the measurement of the force of gravity, realized that if light is affected by the laws of gravity -- as he reasoned it should, given the universal nature of gravity\footnote{\small This is already implicit in {\it Query 1} of Newton's Opticks: {\em ``Do not Bodies act upon Light at a distance, and by their action bend its Rays; and is not this action (c¾teris paribus) strongest at the least distance?"}} -- then there could potentially exist objects that are so massive that even light would not be able to escape their gravitational pull~\cite{1784RSPT...74...35M}. 
%
%so massive could at least in principle exist, that the speed of light would not be sufficient to escape from their gravitational pull, and they therefore would appear dark to a distant observer \cite{1784RSPT...74...35M}. 
%
This proposal, also famously discussed a decade later by Pierre Simon Laplace, is often considered to be the first mention of what have become known as black holes. We mention it here, however, as an explicit example of a discussion of a class of invisible astrophysical objects, that populate the universe while residing beyond the reach of astronomical observations.

% J. Michell Phil. Trans. R. Soc. Lond. 1 January 1784 vol. 74 35-57
% 

The mathematician Friederich Bessel was perhaps the first to predict the existence of a specific undiscovered astronomical object, based only on its gravitational influence. In a letter published in 1844~\cite{1844MNRAS...6R.136B}, he argued that the observed proper motion of the stars Sirius and Procyon could only be explained by the presence of faint companion stars, influencing the observed stars through their gravitational pull:
\begin{quotation}
\em{If we were to regard Procyon and Sirius as double stars, their change of motion would not surprise us.}
\end{quotation} 
\noindent
Bessel further argued in favor of the existence of many stars, possibly an infinite number of them, also anticipating the modern concept of the mass-to-light ratio: 
\begin{quotation}
\em{But light is no real property of mass. The existence of numberless visible stars can prove nothing against the evidence of numberless invisible ones.}
\end{quotation} 

Only two years later, in 1846, the French astronomer Urbain Le Verrier and the English astronomer John Couch Adams, in order to explain some persistent anomalies in the motion of Uranus, proposed the existence of a new planet. Le Verrier's calculations were so precise that the German astronomer John Galle (assisted by Heinrich D'Arrest) identified the new planet at the Berlin observatory the same evening he received the letter from Le Verrier, within 1 degree of the predicted position. 

Interestingly, it was Le Verrier himself who also later noticed the anomalous precession of the perihelion of Mercury, and proposed the existence of a perturbing planet to explain it. As it is well known, this ``dark planet'' -- called {\it Vulcan} -- was never observed, and the solution to this problem would have to await the advent of Einstein's theory of general relativity. 

Beside dark stars and planets, astronomers in the 19th century also discussed dark matter in the form of dark clouds, or dark ``nebulae''. One of the earliest traces of this discussion can be found in a memoir written in 1877 by father Angelo Secchi, then Director of the Roman College Observatory, describing research on nebulae that had been carried out 20 years earlier~\cite{secchi77}:
\begin{quotation}
\em{Among these studies there is the interesting probable discovery of dark masses scattered in space, whose existence was revealed thanks to the bright background on which they are projected. Until now they were classified as black cavities, but this explanation is highly improbable, especially after the discovery of the gaseous nature of the nebular masses.}
\end{quotation} 

Around the end of the 19th century, an interesting discussion began to take place within the astronomical community. As soon as astronomical photography was invented, scientists started to notice that stars were not distributed evenly on the sky. Dark regions were observed in dense stellar fields, and the question arose of whether they were dark because of a paucity of stars, or due to the presence of absorbing matter along the line-of-sight. The astronomer Arthur Ranyard, who was among the main proponents of the latter hypothesis, wrote in 1894~\cite{Ranyard84}: 
%\footnote{Knowledge, 17, 253 (1894)}
\begin{quotation}
\em{The dark vacant areas or channels running north and south, in the neighborhood of [$\theta$ Ophiuchi] at the center .... seem to me to be undoubtedly dark structures, or absorbing masses in space, which cut out the light from the nebulous or stellar region behind them.}
\end{quotation} 

This debate went on for quite some time, and it sparked some interesting ideas. W.~H.~Wesley, who acted for 47 years as the assistant secretary of the Royal Astronomical Society, proposed a novel way to settle the question, involving a rudimentary simulation of the arrangement of stars in the Milky Way~\cite{wesley}:
\begin{quotation}
\em{It is better to solve the question experimentally. For this purpose [the author] repeated many times the experiment of sprinkling small splashes of Indian ink upon paper with a brush, revolving the paper between each sprinkling, so to avoid the chance of showing any artificial grouping in lines due to the direction in which the spots of ink were thrown from the hairs of the brush.}
\end{quotation} 
\subsection{Dynamical Evidence}

%CCCCCCCC Russell, H.N., Dugan, R.S., \& Stewart, J.Q. 1926. Astronomy, Boston: Ginn \& Co. (trimble says they quote Kelvin's argument) CCCCCCCC

Lord Kelvin was among the first to attempt a dynamical estimate of the amount of dark matter in the Milky Way. His argument was simple yet powerful: if stars in the Milky Way can be described as a gas of particles, acting under the influence of gravity, then one can establish a relationship between the size of the system and the velocity dispersion of the stars~\cite{Kelvin04}:

\begin{quotation}
\em{It is nevertheless probable that there may be as many as $10^9$ stars [within a sphere of radius 3.09 $\cdot 10^{16}$ kilometres] but many of them may be extinct and dark, and nine-tenths of them though not all dark may be not bright enough to be seen by us at their actual distances. [...] Many of our stars, perhaps a great majority of them, may be dark bodies. }
\end{quotation} 

Kelvin also obtained an upper limit on the density of matter within such a volume, arguing that larger densities would be in conflict with the observed velocities of stars. Henri Poincar\'e was impressed by Lord Kelvin's idea of applying the ``theory of gases'' to the stellar system of the Milky Way. In 1906 he explicitly mentioned ``dark matter'' (``mati\`ere obscure'' in the original French), and argued that since the velocity dispersion predicted in Kelvin's estimate is of the same order of magnitude as that observed, the amount of dark matter was likely to be less than or similar to that of visible matter \cite{1906Poincare} (for an English translation, see Ref. \cite{1906PA.....14..475P}. See also Ref. \cite{1911lhcp.book.....P} for a more complete discussion):

\begin{quotation}
{There are the stars which we see because they shine; but might there not be obscure stars which circulate in the interstellar space and whose existence might long remain unknown? Very well then, that which Lord Kelvin's method would give us would be the total number of stars including the dark ones; since his number is comparable to that which the telescope gives, then there is no dark matter, or at least not so much as there is of shining matter.}
\end{quotation} 
Along similar lines, in 1915, the Estonian astronomer Ernst \"Opik built a model (published in Russian) of the motion of stars in the Galaxy, also concluding that the presence of large amounts of unseen matter was unlikely~\cite{Opik15}. 

An important step forward in the understanding of the structure of the Milky Way was made by the Dutch astronomer Jacobus Kapteyn. In his most important publication, which appeared shortly before his death in 1922, Kapteyn attempted ``a general theory of the distribution of masses, forces and velocities in the sidereal system" -- that is, in the Milky Way. 

Kapteyn was among the first to offer a quantitative model for the shape and size of the Galaxy, describing it as a flattened distribution of stars, rotating around an axis that points towards the Galactic Pole. He argued that the Sun was located close to the center of the Galaxy, and that the motion of stars could be described as that of a gas in a quiescent atmosphere. He then proceeded to establish a relationship between the motion of stars and their velocity dispersion, similar to what \"Opik had done a few years earlier.

Kapteyn expressed the local density in terms of an effective stellar mass, by dividing the total gravitational mass by the number of observed stars -- including faint ones, through an extrapolation of the luminosity function -- and he explicitly addressed the possible existence of dark matter in the Galaxy:

\begin{quotation}
\em{We therefore have the means of estimating the mass of the dark matter in the universe. As matters stand at present, it appears at once that this mass cannot be excessive. If it were otherwise, the average mass as derived from binary stars would have been very much lower than what has been found for the effective mass.}
\end{quotation} 

In 1932, Kapteyn's pupil, Jan Oort, published an analysis of the vertical kinematics of stars in the solar neighborhood~\cite{1932BAN.....6..249O}. In this work, Oort added to the list of estimates for the local dark matter density, including those by James Jeans (1922)~\cite{1922MNRAS..82..122J} and by Bertil Lindblad (1926)~\cite{Lindblad26}. In his analysis, Oort made a number of improvements on Kapteyn's seminal work, relaxing for instance the assumption of the ``isothermality'' of the gas of stars. 

Oort derived a most probable value for the total density of matter near the Sun of 0.092 $M_{\odot}/$pc$^{3}$, corresponding to $6.3 \times 10^{-24}$ g/cm$^{3}$. He compared this number to the value obtained by Kapteyn, 0.099 $M_{\odot}/$pc$^{3}$, and noticed that the agreement was ``unexpectedly good'', given the differences in treatment and the data used. The numbers obtained by Jeans and Lindblad were each somewhat higher, 0.143 $M_{\odot}/$pc$^{3}$ and 0.217 $M_{\odot}/$pc$^{3}$, respectively.

In order to estimate the amount of dark matter, Oort then proceeded with an estimate for the contribution from stars to the local density, arguing that an extrapolation of the stellar mass function based on the observed stars seemed to be able to account for a substantial fraction of the inferred total density. It is interesting to recall the words used by Oort to illustrate the constraint on the amount of dark matter: 

\begin{quotation}
\em{We may conclude that the total mass of nebulous or meteoric matter near the sun is less than 0.05 M$_{\odot}$pc$^{-3}$, or $3 \cdot 10^{-24}$ g cm$^{-3}$; it is probably less than the total mass of visible stars, possibly much less.}
\end{quotation} 

We learn from this quote not only that the maximum allowed amount of dark matter was about half of the total local density, but also that astronomers thought that the dark matter was likely to consist of faint stars, that could be accounted for through a suitable extrapolation of the stellar mass function, along with ``nebulous'' and ``meteoric'' matter. 

As we shall see in Chapter \ref{Chap:rotcurves} , the pioneering work of Kapteyn, Jeans, Lindblad, \"Opik and Oort opened the path toward modern determinations of the local dark matter density, a subject that remains of importance today, especially for experiments that seek to detect dark matter particles through their scattering with nuclei. 

%While our understanding of the dynamics of our Galaxy was still in its infancy, and continued to be hotly debated, a Swiss astronomer working at Caltech, obtained the first strong indication of the presence of large quantities of dark matter in the Coma cluster o. Fritz Zwicky's groundbreaking work, its implications, and the works that followed are the subject of next chapter.

%\section*{Further reading}
%Trimble 2013, \cite{2013pss5.book.1091T}
%\bibitem[Bessel(1844)]{1844MNRAS...6R.136B} Bessel, F.~W.\ 1844, \mnras, 6, 136 
% Virginia Trimble
% Jan Oort, including his new book
% Sanders book 
% Kragh

\section{Galaxy Clusters}
\label{sec:clusters}

\subsection{Zwicky and Smith}
\label{zwickysmith}

The Swiss-American astronomer Fritz Zwicky is arguably the most famous and widely cited pioneer in the field of dark matter. In 1933, he studied the redshifts of various galaxy clusters, as published by Edwin Hubble and Milton Humason in 1931~\cite{1931ApJ....74...43H}, and noticed a large scatter in the apparent velocities of eight galaxies within the Coma Cluster, with differences that exceeded 2000 km/s~\cite{1933AcHPh...6..110Z}. The fact that Coma exhibited a large velocity dispersion with respect to other clusters had already been noticed by Hubble and Humason, but Zwicky went a step further, applying the virial theorem to the cluster in order to estimate its mass. 

This was not the first time that the virial theorem, borrowed from thermodynamics, was applied to astronomy; Poincare had done so more than 20 years earlier in his {\it Le\c{c}ons sur les hypoth\`{e}ses cosmogoniques profess\'{e}es \`a la Sorbonne} \cite{1911lhcp.book.....P}. But to the best of our knowledge, Zwicky was the first to use the virial theorem to determine the mass of a galaxy cluster.

Zwicky started by estimating the total mass of Coma to be the product of the number of observed galaxies, 800, and the average mass of a galaxy, which he took to be $10^9$ solar masses, as suggested by Hubble. He then adopted an estimate for the physical size of the system, which he took to be around $10^6$ light-years, in order to determine the potential energy of the system. From there, he calculated the average kinetic energy and finally a velocity dispersion. He found that 800 galaxies of $10^9$ solar masses in a sphere of $10^6$ light-years should exhibit a velocity dispersion of 80 km/s. In contrast, the observed average velocity dispersion along the line-of-sight was approximately 1000 km/s. From this comparison, he concluded: 
\begin{quotation}
\emph{If this would be confirmed, we would get the surprising result that dark matter is present in much greater amount than luminous matter.}
\end{quotation} 
\noindent

This sentence is sometimes cited in the literature as the first usage of the phrase ``dark matter''. It is not, as we have seen in the previous chapter, and it is not even the first time that Zwicky used it in a publication. He had, in fact, used the same phrase in a article published earlier the same year, pertaining to the sources of cosmic rays~\cite{PhysRev.43.147}:

\begin{quotation}
\emph{According to the present estimates the average density of dark matter in our galaxy $(\rho_g)$ and throughout the rest of the universe $(\rho_u)$ are in the ratio $\rho_g / \rho_u >$ 100,000.}
\end{quotation} 
\noindent
Although he doesn't explicitly cite any article, it is obvious from this sentence that he was well aware of the work of Kapteyn, Oort and Jeans discussed in the previous chapter. His use of the term ``dark matter'' is, therefore, in continuity with the community of astronomers that had been studying the dynamics of stars in the local Milky Way.

In 1937, Zwicky published a new article -- this time in English, in {\it the Astrophysical Journal}~\cite{1937ApJ....86..217Z} -- in which he refined and extended his analysis of the Coma Cluster. The purpose of this paper was to determine the mass of galaxies, and he proposed a variety of methods to attack this problem. In particular, he returned to the virial theorem approach that he had proposed in 1933, this time assuming that Coma contained 1000 galaxies within a radius of $2\times 10^6$ light-years, and solving for the average galaxy's mass. From the observed velocity dispersion of 700 km/s, he obtained a conservative lower limit of $4.5 \times 10^{13}\, M_{\odot}$ on the mass of the cluster  (to be conservative, he excluded a galaxy with a recession velocity of 5100 km/s as a possible outlier), corresponding to an average mass-per-galaxy of $4.5 \times 10^{10} \, M_{\odot}$. Assuming then an average absolute luminosity for cluster galaxies of $8.5 \times 10^7$ times that of the Sun, Zwicky showed that this led to a surprisingly high mass-to-light ratio of about 500. 

Zwicky's work relied on Hubble's relationship between redshift and distance, and in the 1937 paper he used the results of Hubble and Humason~\cite{1931ApJ....74...43H}, which pointed to a Hubble constant of $H_0 = 558$ km/s/Mpc, with an estimated uncertainty of 10-20\%. If we rescale these results adopting the modern value of $H_0 = 67.27 \pm$ 0.66~\cite{2015arXiv150201589P}, we see that Zwicky overestimated the mass-to-light ratio by a factor of $\sim 558 / 67.27 =  8.3$. Despite this substantial correction, Coma's velocity dispersion still implies a very high mass-to-light ratio and points to the existence of dark matter in some form.

%The mass-to-light ratio obtained by rescaling Zwicky's estimate by this factor is $500/60=60$, which is still very high and points to the existence of some form of dark matter. 

What did Zwicky think that the dark matter in Coma and other galaxy clusters might be? An illuminating sentence in his 1937 paper provides a rather clear answer to this question: 
\begin{quotation}
\emph{[In order to derive the mass of galaxies from their luminosity] we must know how much dark matter is incorporated in nebulae in the form of cool and cold stars, macroscopic and microscopic solid bodies, and gases.}
\end{quotation} 
\noindent

Meanwhile, another estimate for the mass of a cluster of galaxies had appeared in 1936, this time from Sinclair Smith, who had studied the Virgo Cluster. Assuming that the outer galaxies were in circular motion around Virgo, Smith calculated a total mass for the cluster of $10^{14}\, M_{\odot}$. When divided by the number of observed galaxies, 500, he found an average mass per galaxy of $2 \times 10^{11}\, M_{\odot}$, which he pointed out was much higher than Hubble's estimate of $10^9$ M$_{\odot}$.

Much like Zwicky, whose 1933 work he cites, Smith considers this high value for the mass-per-galaxy implied by his calculations to be a problem, in particular in light of its incompatibility with Hubble's estimate. He also acknowledges, however, that both could be correct, and that: 
\begin{quotation}
\emph{the difference represents internebular material, either uniformly distributed or in the form of great clouds of low luminosity surrounding the [galaxies].}
\end{quotation} 
\noindent

In his famous book {\it The Realm of Nebulae}, Hubble cites the work of Smith (and not that of Zwicky), and clearly states that he considers the discrepancy between the masses of galaxies inferred from the dynamics of clusters and those from the rotation of galaxies to be ``real and important''. And although he argued that this problem might be solved, or at least diminished, by observing that the former were likely upper limits, while the latter lower limits, he acknowledged that this argument was not entirely satisfactory. A confusing situation had indeed arisen. 

\subsection{A Confusing Situation}

There was no shortage of reasons for astronomers to be skeptical of the findings of Zwicky and Smith. The assumption that Virgo was a system in equilibrium, made by Smith, was questioned by Zwicky himself in his 1937 paper. In 1940, Erik Holmberg -- who will appear again in this review as a pioneer of numerical simulations -- described some of the concerns of the community regarding the work of Zwicky and Smith~\cite{1940ApJ....92..200H}:
\begin{quotation}
\emph{It does not seem to be possible to accept the high velocities [in the Virgo and Coma cluster] as belonging to permanent cluster members, unless we suppose that a great amount of mass -- the greater part of the total mass of the cluster -- is contributed by dark material distributed among the cluster members -- an unlikely assumption.}
\end{quotation} 
\noindent

\begin{figure}
    \begin{center}
        \includegraphics[width=\textwidth]{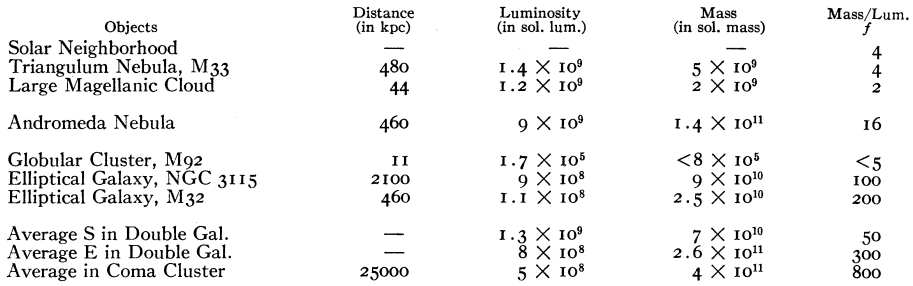}
    \caption{A snapshot of the dark matter problem in the 1950s: the distance, mass, luminosity, and mass-to-light ratio of several galaxies and clusters of galaxies, as compiled by M.~Schwarzschild in 1954~\cite{1954AJ.....59..273S}.}
    \label{fig:Schw54}
    \end{center}
\end{figure}

Holmberg argued instead that these galaxies were probably ``temporary'' members of the cluster, i.e.~galaxies on hyperbolic orbits that had fallen into the gravitational potential of the cluster, but were not bound to it. In 1954, Martin Schwarzschild~\cite{1954AJ.....59..273S} -- son of the famous Karl Schwarzchild who had made important contributions to general relativity -- attempted to get rid of ``interlopers'', and inferred a smaller radial velocity dispersion of 630 km/s. By adopting an updated Hubble parameter, and an average luminosity-per-galaxy of $5\times 10^8\,L_{\odot}$, he obtained the ``bewildering high'' mass-to-light ratio of 800. The distance, mass, luminosity, and mass-to-light ratio of the galaxies and clusters of galaxies compiled by Schwarzschild are shown in Fig.~\ref{fig:Schw54}.

By the late 1950s, a number of other articles had been published on the mass-to-light ratios of galaxy clusters. Victor Ambartsumian rejected the possibility that dark matter existed in clusters and argued instead that they are unstable and rapidly expanding systems, to which the virial theorem cannot be applied. It was soon realised, however, (e.g.~Burbidge and Burbidge~\cite{1959ApJ...130..629B} and Limber~\cite {1962IAUS...15..239L}) that this interpretation was in tension with the estimated age of the galaxies (requiring clusters that were younger than the galaxies they contained), and with that of the Universe (the clusters should have evaporated long ago). In August of 1961, a conference on the instability of systems of galaxies was held in Santa Barbara, and included as participants some of the most important astrophysicists active in that field of research. Jerzy Neyman, Thornton Page and Elizabeth Scott summarized the discussions that took place around the mass discrepancy in galaxy clusters as follows: 
\begin{quotation}
\emph{Several possible explanations of this mass discrepancy were discussed at the Conference [..]. Many of those present consider that it might be real and due to invisible inter-galactic material in the clusters, totalling 90 to 99\% of their mass. If these possibilities are excluded, however, the discrepancy in mass indicates positive total energy and instability of the system involved.}
\end{quotation} 
\noindent

%\begin{figure}
%    \begin{center}
%        \includegraphics[width=\textwidth]{Limber.png}
%    \caption{A schematic representation of possible explanations for the mass discrepancy in galaxy clusters, circa 1962~\cite{1954AJ.....59..273S}.}
%    \label{fig:Limber}
%    \end{center}
%\end{figure}

The overall situation was that of a community that was struggling to find a unified solution to a variety of problems. The dark matter hypothesis was not commonly accepted, nor was it disregarded. Instead, there was a consensus that more information would be needed in order to understand these systems. 

%The new information that was needed did begin to appear in the late 1960s, in the form of evidence for large quantities of dark matter in galaxies (see the following chapter). 
 
In addition to the question of whether the dynamics of galaxy clusters required the presence of dark matter, astronomers around this time began to be increasingly willing to contemplate what this dark matter might be made of. Herbert Rood~\cite{1965PhDT.........3R} (later confirmed by Simon White~\cite{1977MNRAS.179...33W}) studied the relaxation process of galaxy clusters and argued that the mass responsible for their high mass-to-light ratios must to be found within the intergalactic space, and not in the galaxies themselves. 
%But where was this material, and how could it be searched for? 
Arno Penzias searched for free hydrogen in the Pegasus I cluster and set an upper limit of a tenth of its virial mass \cite{1961AJ.....66R.293P}. Neville Woolf suggested in 1967 that the gas could be ionised, and used radio, visible and X-ray observations to set limits on it~\cite{1967ApJ...148..287W}. Turnrose and Rood discussed the problems of this hypothesis in Ref.~\cite{1970ApJ...159..773T}, and in 1971 Meekins et al.~\cite{1971Natur.231..107M} obtained observational evidence for X-ray emission that limited the amount of hot intracluster gas to be less than 2\% of that required for gravitational binding.

With gas ruled out as an explanation for the ``missing mass'' in galaxy clusters, scientists began to explore more or less exotic possibilities, including massive collapsed objects~\cite{1969Natur.224..891V}, HI snowballs~\cite{1971phco.book.....P}, and M8 dwarf stars~\cite{1974QJRAS..15..122T}. As we will discuss in Chapter~V, these possibilities -- and others like them -- were eventually ruled out by measurements of the primordial light element abundances, which instead favor a non-baryonic nature for the dark matter.

%\subsection*{Further reading}:
%
%\begin{itemize}
%\item Title:	 From Messier to Abell: 200 Years of Science with Galaxy Clusters
%
%\end{itemize}

\section{Galactic Rotation Curves}
\label{Chap:rotcurves}

\subsection{The Beginnings}

The rotation curves of galaxies --  i.e. the circular velocity profile of the stars and gas in a galaxy, as a function of their distance from the galactic center -- played a particularly important role in the discovery of dark matter. Under some reasonable simplifying assumptions, it is possible to infer the mass distribution of galaxies from their rotation curves. Historically, it was the observation of approximately ``flat'' rotation curves at very large galactocentric distances that did the most to convince the scientific community that large amounts of dark matter is present in the outer regions of galaxies. 

% convincingly proved the existence of large amounts of dark matter in the outer parts of galaxies.

In 1914, ten years before Hubble convincingly demonstrated that Andromeda (M31) was a galaxy and located outside of the Milky Way, Max Wolf~\cite{wolf14} and Vesto Slipher~\cite{1914LowOB...2...66S} noticed that the spectral lines from this system were inclined when the slit of the spectrogram was aligned with the galaxy's major axis and straight when it was aligned with the minor axis, allowing them to conclude that Andromeda rotates. Based on 79 hours of observation in 1917 with the Mount Wilson Observatory's 60-inch reflector, Francis Pease measured the rotation of the central region of Andromeda out to an angular radius of 2.5 arcminutes, finding that it rotates with an approximately constant angular velocity.  Several authors used Andromeda's observed rotational velocity to calculate its mass and discuss its mass-to-light ratio in comparison with the measured value for the solar neighborhood (see Chapter~\ref{prehistorychapter}), finding values that were in reasonable agreement, e.g.~Hubble (1926) \cite{1926ApJ....64..321H}, Oort (1932) \cite{1932BAN.....6..249O}. 

In a paper published in 1930~\cite{Lundmark30}, Knut Lundmark made estimates for the mass-to-light ratios of five galaxies based on a comparison of their absolute luminosity -- as estimated using novae as distance indicators -- and their mass as inferred from spectroscopic observations. These mass-to-light ratios varied, quite unrealistically, from 100 for M81 to 6 for M33 -- much larger than those found for the solar neighborhood (Lundmark also made early estimates for the mass of the Milky Way~\cite{1925MNRAS..85..865L}). This demonstrates that astronomers at the time were open to the possibility that large amounts of dark matter might be present in astrophysical systems, in the form of ``extinguished stars, dark clouds,  meteors, comets, and so on'', as Lundmark writes in 1930\footnote{Translation from the German by Lars Bergstr{\"o}m}. Holmberg argued in 1937 that the large spread in mass-to-light ratios found by Lundmark was a consequence of the absorption of light {\it ``produced by the dark matter''}, and that once this was correctly taken into account, all of the galaxies studied by Lundmark, including the Milky Way, would have mass-to-light ratios between 6 and 7 \cite{1937AnLun...6....3H}. 

Fritz Zwicky, in his famous 1937 article on galaxy clusters, discussed the possibility of using the rotation curves of galaxies to infer their mass distribution, concluding that:
\begin{quotation}
\emph{It is not possible to derive the masses of [galaxies] from observed rotations, without the use of additional information.}
\end{quotation} 
\noindent
Beside the lack of information on the ellipticity of orbits, one of Zwicky's main concerns was the possible internal ``viscosity'' resulting from the mutual interactions of stars. Only four years later, Chandrasekhar would demonstrate in his classic paper, ``The Time of Relaxation of Stellar Systems'', that these interactions are completely negligible, allowing one to reliably describe galaxies as systems of non-interacting stars.

Meanwhile, in his 1939 PhD dissertation, Horace Babcock presented the rotation curve of M31 out to 100 arcminutes (i.e.~about 20 kpc) away from its center~\cite{1939LicOB..19...41B}.  Interestingly, he found very high values for the circular velocity at large radii -- so high, in fact, that they are at odds with modern measurements. Approximating M31 as a sphere surrounded by a flattened ellipsoid, he calculated the mass distribution of the galaxy, recognizing that the observed rising rotation curve at large radii implied the existence of large amounts of mass in the outer parts of the galaxy. But when interpreting this result, he conservatively argued that: 
\begin{quotation}
\emph{the calculated ratio of mass to luminosity in proceeding outward from the nucleus suggests that absorption plays a very important role in the outer portion of the spiral, or, perhaps, that new dynamical considerations are required, which will permit of a smaller mass in the outer parts.}
\end{quotation} 
\noindent

More than a decade later, observations made by Nicholas Mayall in 1951 at Mount Wilson~\cite{1954AJ.....59..273S} were used by Martin Schwarzschild to further study the dynamics of M31. In doing so, Schwarzchild showed that a model with a constant mass-to-light ratio was able to explain the rotational velocities measured by Mayall out to 115 arcminutes.  

%\section{The Advent of Radio Astronomy}
The German invasion of Poland in 1939 marked the official start of World War II. Hostilities brought death and destruction, but also unexpected benefits for science, as after the war ended military radars began to be used for radio astronomical observations. The Netherlands was particularly active in this field, under the push of the visionary astronomer Jan Oort, who was not only a great scientist, but also a great organizer. A chain of so-called W\"urzburg antennas -- 7.5 meter parabolic radars used at 54 cm wavelengths for aircraft tracking -- had been left behind in the Netherlands by occupying German forces at the end of the war, and since the reflective surface and tracking precision were also suitable for shorter wavelengths, and in particular for the 21 cm line predicted by Oort's student Hendrik van de Hulst, one was mounted in Kootwijk for the purpose of radio astronomy~\cite{2006JAHH....9....3V}. 

When Harold Ewen and Edward Purcell, from Harvard, detected the 21 cm line in 1951, van de Hulst was visiting Harvard, and so was F.~J.~Kerr from the Radiophysics Laboratory in Sydney. The Dutch and Australian groups were soon able to confirm the detection: the reports of the American and Dutch groups appeared in the same issue of {\it Nature}, together with a confirming telegram from the Australian group. This success provided an important boost to the young field of radio astronomy, and had a dramatic impact on the history of astrophysics and cosmology.

Back in the Netherlands, the construction of a new 25 meter radio telescope was completed in Dwingeloo, in 1955. Only two years later, van de Hulst, Jean Jacques Raimond, and Hugo van Woerden published the first radio rotation curve of M31, extending observations to 2 degrees away from its center ~\cite{1957BAN....14....1V}. Although the data seemed at first to be at odds with the rotation curve calculated by Schwarzschild, Maartin Schmidt argued in a paper accompanying the publication of van de Hulst et al.~that a constant mass-to-light ratio provided a satisfactory explanation of the data, although also noting that ``nothing as yet can be stated about the ratio in the innermost and outermost parts'' of M31~\cite{1957BAN....14...17S}.

In 1959, Franz Kahn and Lodewijk Woltjer proposed an ingenious method to determine the combined mass of M31 and the Milky Way. Since 21 cm observations of M31 indicated that it was approaching the Milky Way at a speed of 125 km/s, they derived a lower bound on the reduced mass of the M31-Milky Way system, assuming that the two galaxies are part of a bound system and that the orbital period is smaller than the age of the Universe. That lower bound was, however, six times larger than the currently accepted value of the reduced mass of the system \cite{1959ApJ...130..705K}. The authors argued at the time that this provided evidence for intergalactic material in the form of gas stabilising the local group. In retrospect, this simple argument is one of the earliest clear indications of dark matter halos around galaxies. 

In his detailed historical account~\cite{2010dmp..book.....S}, Robert Sanders argues that despite these developments there was no sense of crisis in the field of astrophysics at the end of the 1950s, or at least that there was no consensus that the observed rotation curves were in conflict with the current understanding of galaxies. A decade later, things began to dramatically change. 

\subsection{The 1970s Revolution}

In the 1960s, Kent Ford developed an image tube spectrograph that Vera Rubin and he used to perform spectroscopic observations of the Andromeda Galaxy. 
The observations of the M31 rotation curve Rubin and Ford published in 1970 \cite{1970ApJ...159..379R} represented a step forward in terms of quality. Their optical data extended out to 110 arcminutes away from the galaxy's center, and were compatible with the radio measurements obtained previously by Morton Roberts in 1966~\cite{1966ApJ...144..639R}.

\begin{figure}[ht]
    \begin{center}
        \includegraphics[width=0.6\textwidth]{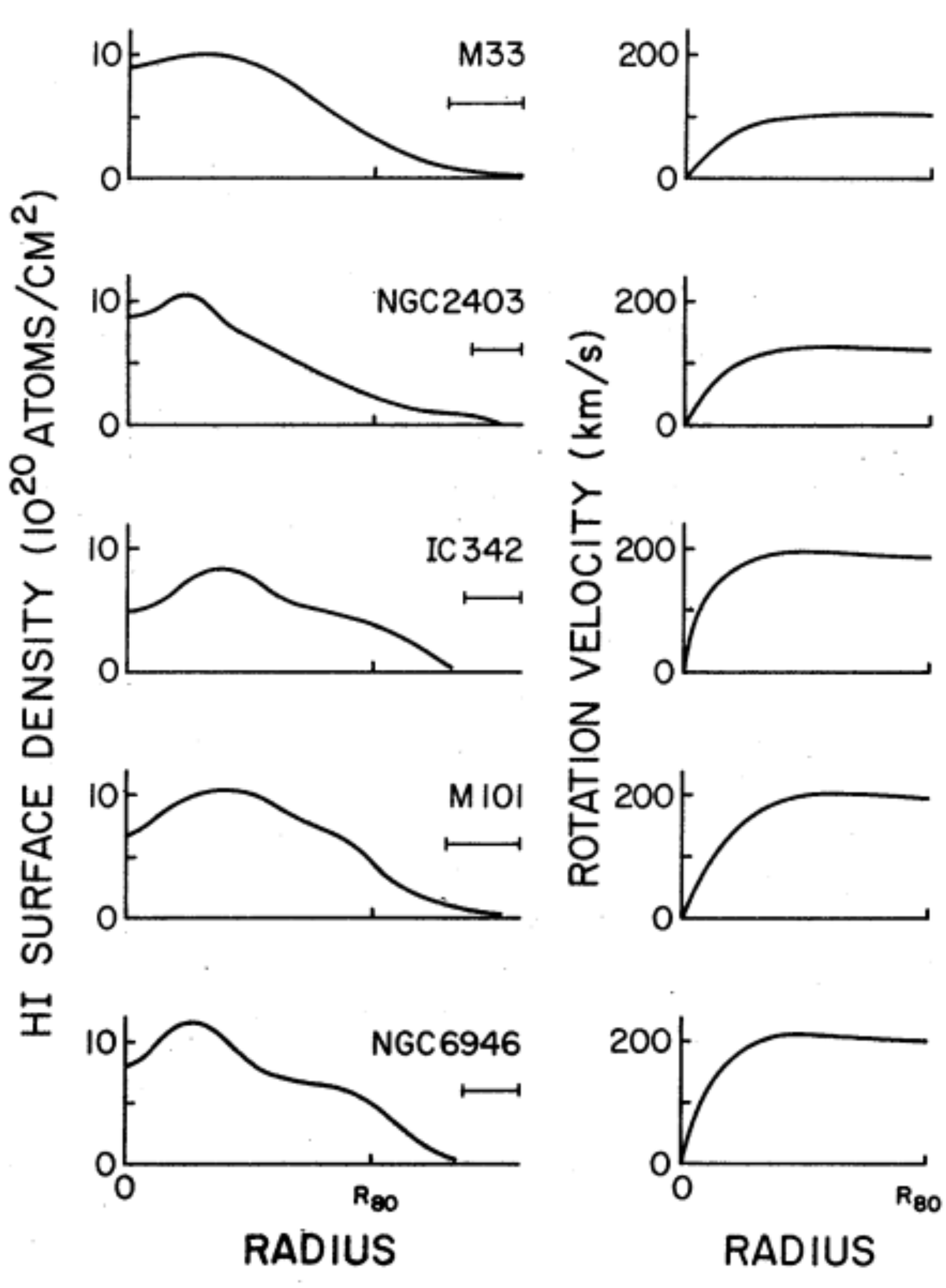}
    \caption{Flat rotation curves began to emerge clearly from 21 cm observations in the early 1970s. Here we show the hydrogen surface density profile (left) and the rotation curves (right) of five galaxies as obtained by Rogstad and Shostak in 1972~\cite{1972ApJ...176..315R}. The bars under the galaxy names indicate the average radial beam diameter, i.e.~the effective spatial resolution. R80 is the radius containing 80\% of the observed HI.}
    \label{fig:flat72}
    \end{center}
\end{figure}

It was also in 1970 that the first explicit statements began to appear arguing that additional mass was needed in the outer parts of some galaxies, based on comparisons of the rotation curves predicted from photometry and those measured from 21 cm observations. In the appendix of his seminal 1970 paper~\cite{1970ApJ...160..811F}, Ken Freeman compared the radius at which the rotation curve was predicted to peak, under the assumption of an exponential disk with a scale length fit to photometric observations, to the observed 21 cm rotation curve. This combination of theoretical modelling and radio observations extending beyond the optical disk allowed Freeman to reach a striking conclusion. He found that for M33 (based on data summarised in Ref.~\cite{1965MNRAS.129..309B}) and NGC 300 (based on data from Ref.~\cite{1967AuJPh..20..131S}), the observed rotation curves peaked at larger radii than predicted, and -- prompted by discussions with Roberts\footnote{K. Freeman, private communication.} -- concluded that:
\begin{quotation}
\emph{if [the data] are correct, then there must be in these galaxies additional matter which is undetected, either optically or at 21 cm. Its mass must be at least as large as the mass of the detected galaxy, and its distribution must be quite different from the exponential distribution which holds for the optical galaxy.}
\end{quotation} 
\noindent
This is perhaps the first convincing (or at least convinced) claim of a mass discrepancy in galaxies. D.~Rogstad and G.~Shostak performed a similar analysis in 1972 \cite{1972ApJ...176..315R}, by analyzing the rotation curves of five galaxies -- M33, NGC 2403, IC 342, M101 and NGC 6946 --  they had themselves obtained using the radio telescope at the Owens Valley Radio Observatory. They found that these rotation curves remained flat out to the largest radii observed (see Fig.~\ref{fig:flat72}) and, following the method of Freeman, they derived mass-to-light ratios as high as 20 at large radii. As explicitly said in their paper, they:
\begin{quotation}
\emph{confirm[ed] the requirement of low-luminosity material in the outer regions of these galaxies.}
\end{quotation} 
\noindent

\begin{figure}
    \begin{center}
        \includegraphics[width=0.7\textwidth]{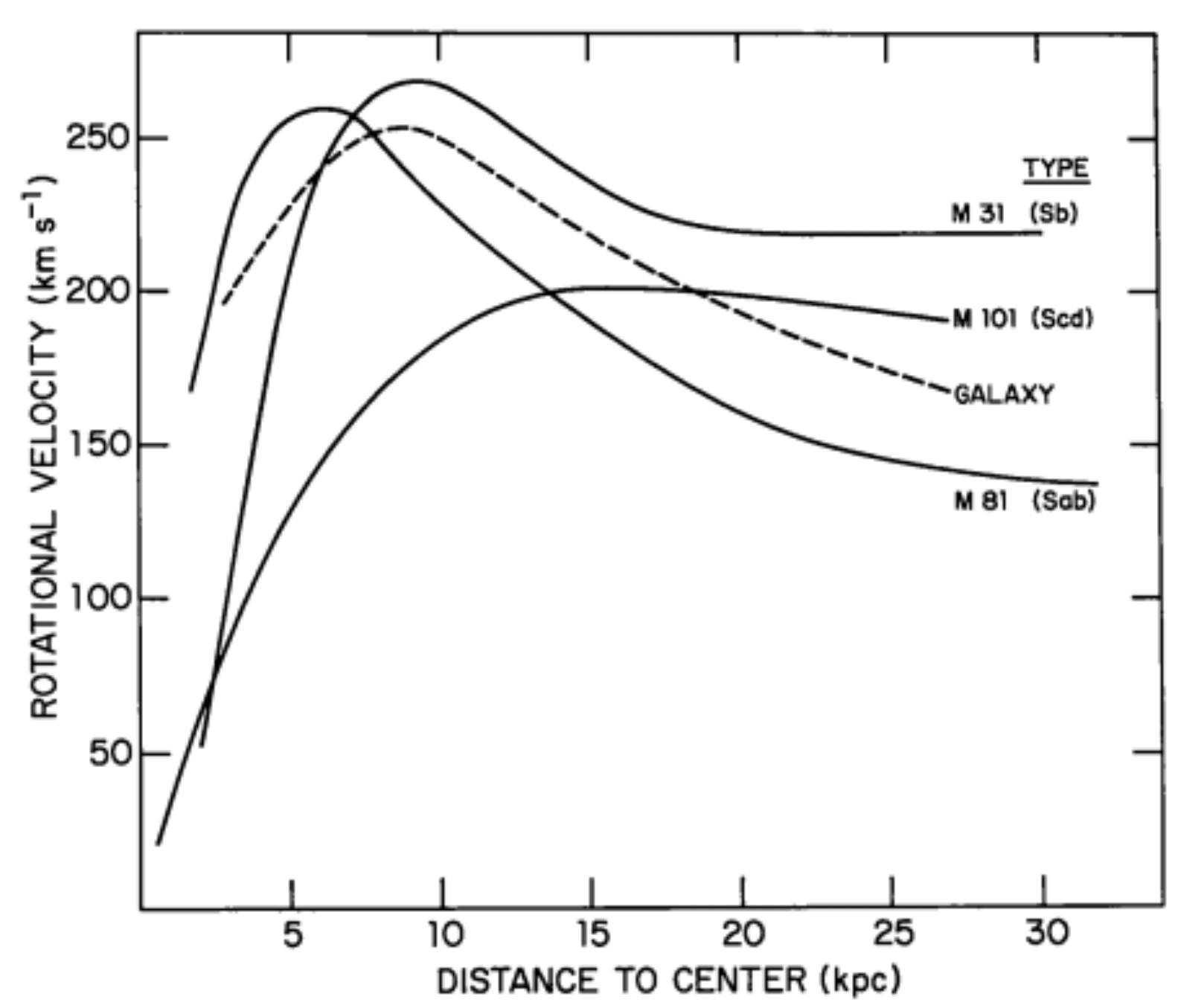}
    \caption{The rotation curves for the galaxies M31, M101, and M81 (solid lines) obtained by Roberts and Rots in 1973. The rotation curve of the Milky Way Galaxy was included by the authors for comparison. From Ref.~\cite{1973A&A....26..483R}.}
    \label{fig:robertsrots1973}
    \end{center}
\end{figure}

Morton Roberts was among the first to recognize the implications of the observed flatness of galactic rotation curves. Together with R.~Whitehurst, he published in 1972 a rotation curve of M31 that extended to 120 arcminutes from its center~\cite{1972ApJ...175..347W}. In 1973, together with Arnold Rots, he extended the analysis to M81 and M101, and argued that these spiral galaxies each exhibited flat rotation curves in their outer parts~\cite{1973A&A....26..483R} (see Fig.~\ref{fig:robertsrots1973}). The authors' interpretation of these data was unambiguous: 
\begin{quotation}
\emph{The three galaxies rotation curves decline slowly, if at all, at large radii, implying a significant mass density at these large distances. It is unreasonable to expect the last measured point to refer to the `edge' of the galaxy, and we must conclude that spiral galaxies must be larger than indicated by the usual photometric measurements [...]. The present data also require that the mass to luminosity ratio vary with radius increasing in distance from the center.}
\end{quotation} 
\noindent

In the Proceedings of the IAU Symposium No.~69, held in Besan\c{c}on, France in September of 1974, Roberts reviewed the status of galactic rotation curves~\cite{1975IAUS...69..331R}, highlighting the importance of radio observations, which extended well beyond the optical radius of the galaxies. When discussing the implications of the high mass-to-light ratios implied by these observations, he argued that the excess mass might take the form of intermediate and late dwarf M stars. He further tried to reassure his colleagues by arguing that the required radius-dependent luminosity function need not be alarming, since there was evidence of a dependence on the height above the Galactic Plane that exhibited a similar trend.

\begin{figure}[t]
    \begin{center}
        \includegraphics[width=0.9\textwidth]{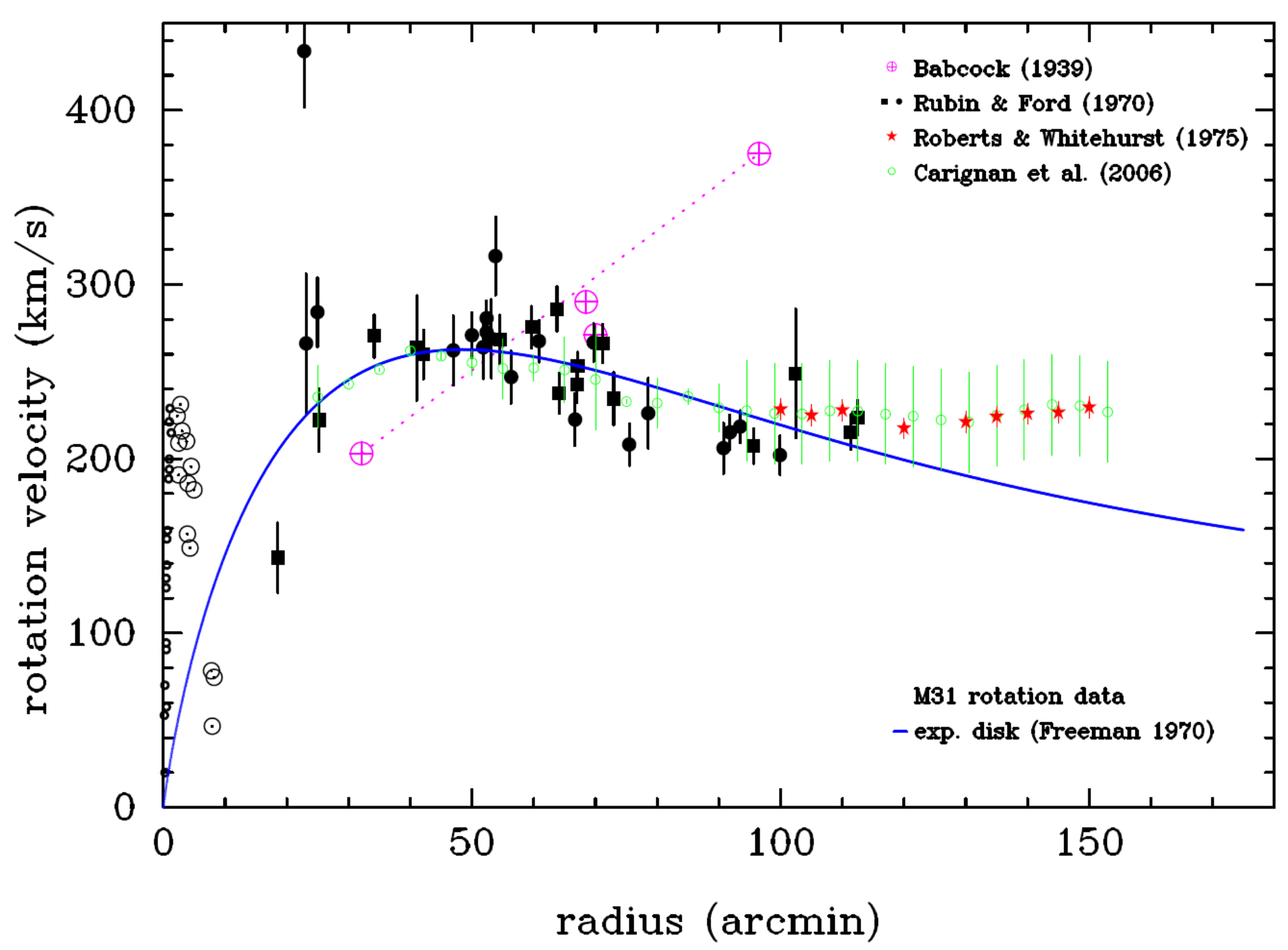}
    \caption{Rotation curve data for M31. The purple points are emission line data in the outer parts from Babcock 1939 \cite{1939LicOB..19...41B}. The black points are from Rubin and Ford 1970 \cite{1970ApJ...159..379R} (squares for the SW data, filled circles for the NE data, and open circles for the data in the inner parts -- the presence of  non-circular motions in the inner parts makes the modelling of those data uncertain). The red points are the 21-cm HI line data from Roberts and Whitehurst 1975 \cite{1975ApJ...201..327R}. The green points are 21-cm HI line data from Carignan et al. \cite{2006ApJ...641L.109C}. The black solid line corresponds to the rotation curve of an exponential disc with a scalelength according to the value given in Freeman 1970 \cite{1970ApJ...160..811F}, suitably scaled in velocity. 21-cm data demonstrate clearly the mass discrepancy in the outer parts. Figure courtesy of Albert Bosma.}
    \label{fig:m31}
    \end{center}
\end{figure}

As we will discuss in Chapter~\ref{chap:cosmology}, two influential papers in 1974 brought together the observed mass discrepancies observed in clusters and in galaxies~\cite{1974Natur.250..309E,1974ApJ...193L...1O}. Both of these papers clearly state in their first paragraph that the mass of galaxies had been until then underestimated by about a factor of ten. In support of this, Jerry Ostriker, Jim Peebles, and Amos~Yahil~\cite{1974ApJ...193L...1O} cited Roberts and Rots~\cite{1973A&A....26..483R} and Rogstad and Shostak~\cite{1972ApJ...176..315R} for the observed flat rotation curves. For the same purpose, Jaan Einasto, Ants Kaasik and Enn Saar~\cite{1974Natur.250..309E} cited a review written in 1975 by Roberts for a book edited by A.~and M.~Sandage together with J.~Cristian~\cite{1975gaun.book.....S}. In a separate paper that appeared in the same year, focusing on the ``morphological evidence'' of missing mass around galaxies, Einasto and collaborators cited the 1973 paper of Roberts and Rots~\cite{1973A&A....26..483R}. Interestingly, Einasto and collaborators excluded the possibility that this missing mass was in the form of stars, and argued that the most likely explanation was the presence of large amounts of gas in the outer parts of galaxies, which they referred to as ``coronas''~\cite{1974Natur.252..111E}.

By 1974, the flat rotation curves obtained by radio astronomers had done much to establish the existence of large amounts of mass in the outer parts of galaxies -- at least to the eyes of the influential authors of Refs.~\cite{1974Natur.250..309E,1974ApJ...193L...1O}. Portions of the astronomical community, however, were still not convinced of this conclusion~\cite{Rubin04}. In the late 1970s, this evidence was strengthened and corroborated by a series of new studies. In 1977, Nathan Krumm and Edwin Salpeter~\cite{1977A&A....56..465K} observed six spiral galaxies with the Arecibo Observatory, and showed that they each exhibited a flat rotation curve out to radii larger than their optical extent, but these data turned out to be unreliable due to beam-smearing (see the discussion at the end of Ref. \cite{1978IAUS...77...23S}). 

\begin{figure}[t]
    \begin{center}
        \includegraphics[width=0.65\textwidth]{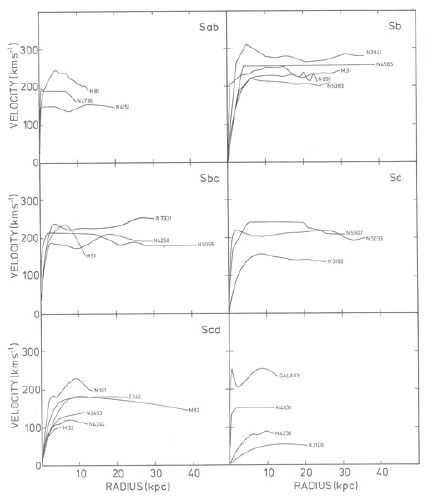}
    \caption{The rotation curves of the 25 galaxies published by Albert Bosma in 1978~\cite{1978PhDT.......195B}.}
    \label{fig:bosma}
    \end{center}
\end{figure}

In 1978, Albert Bosma published the results of his PhD thesis~\cite{1978PhDT.......195B}, including the radio observation of the velocity fields and corresponding rotation curves of 25 galaxies. This work convincingly proved that most of these objects had flat rotation curves out to the largest observed radius, which again exceeded the optical size of the galaxies, therefore demonstrating that their mass continued to grow beyond the region occupied by the stars and gas (see also Fig.~\ref{fig:bosma}). 

A few months later, Rubin, Ford and Norbert Thonnard published optical rotation curves for ten high-luminosity spiral galaxies and found that they were flat out to the outermost measured radius~\cite{1978ApJ...225L.107R}. This work has become one of the most well-known and widely cited in the literature, despite the fact that the optical measurements did not extend to radii as large as those probed by radio observations, thus leaving open the possibility that galaxies may not have dark matter halos, as pointed out, for example, by Agris J. Kalnajs in 1983 (see the discussion at the end of Ref.~\cite{1983IAUS..100...81H}) and by Stephen Kent in 1986~\cite{1986AJ.....91.1301K}. Rubin, Ford and Thonnard themselves acknowledged the credit that was due to the preceding analyses:
\begin{quotation}
\emph{Roberts and his collaborators deserve credit for first calling attention to flat rotation curves. [...]  These results take on added importance in conjunction with the suggestion of Einasto, Kaasik, and Saar (1974) and Ostriker, Peebles and Yahil (1974) that galaxies contain massive halos extending to large $r$.}
\end{quotation} 
\noindent

A lucid and timely review of the status of galaxy masses and mass-to-light-ratios appeared in 1979, authored by Sandra Faber and John Gallagher~\cite{1979ARA&A..17..135F}. We refer the reader to this excellent article for an overview of the various ideas that had been put forward in an effort to understand the complex and diverse observational data that was available at the time. The abstract of that article provides a clear indication of its contents: 
\begin{quotation}
\emph{The current status of the `missing mass' problem is reviewed on the basis of standardized mass-to-light (M/L) ratios of galaxies. The stellar mass density in the immediate vicinity of the sun is examined, along with the mass of the Milky Way and the M/L ratios of spiral galaxies, E and S0 galaxies, and binary galaxies. The dynamics of small groups of galaxies is investigated, and mass derivations for cluster galaxies are discussed. It is concluded that the case for invisible mass in the universe is very strong and becoming stronger.}
\end{quotation} 
\noindent

\subsection{Local Measurements}

We conclude this chapter with a brief overview of the efforts to determine the local dark matter density, i.e.~the density of dark matter in the solar neighborhood. This quantity was historically important, as it provided the first -- albeit rather weak -- dynamical evidence for matter in the local Universe beyond visible stars. It is also important today, as the prospects for detecting dark matter particles in underground and astrophysical experiments strongly depend on this quantity.

As we have seen in the previous chapter, Kapteyn, Lindblad, Jeans and Oort had studied the dynamics of nearby stars, and compared the inferred gravitational mass with that of the visible stellar density. After decades of steady improvements  (Oort 1932 \cite{1932BAN.....6..249O}, Hill (1960), Oort 1960 \cite{1960BAN....15...45O}, Bahcall 1984 \cite{1984ApJ...287..926B,1984ApJ...276..169B}), Konrad Kuijken and Gerry Gilmore published a series of papers based on a refined method and a volume complete sample of K-dwarf data, to derive a much more precise value of the local density~\cite{1989MNRAS.239..605K}. 
The advent of the Hipparcos, SDSS, and RAVE surveys has more recently triggered many new analyses. We refer the reader to the excellent review by Justin Read~\cite{2014JPhG...41f3101R} for further details and references. 

Alternatively, the local dark matter density can be constrained using measurements of the Milky Way's rotation curve (e.g.~Fich et al.~1989~\cite{1989ApJ...342..272F}, Merrifield 1992~\cite{1992AJ....103.1552M}, Dehnen and Binney 1998~\cite{1998MNRAS.294..429D}, Sofue et al.~2009~\cite{2009PASJ...61..227S}, Weber and de Boer 2010~\cite{2010A&A...509A..25W}, Catena and Ullio~2010 \cite{2010JCAP...08..004C}, Salucci et al.~2010~\cite{2010A&A...523A..83S}, Iocco et al.~2011~\cite{Iocco:2011jz}, Pato et al.~2015~\cite{Pato:2015dua}). Although rather precise determinations can be made using this approach, the results strongly depend on the assumptions one makes about the shape of the halo.  Upcoming astronomical surveys -- and in particular the Gaia satellite -- are expected to lead to significant improvements in the reconstruction of the local density (Perryman et al.~2001~\cite{2001A&A...369..339P}, Wilkinson et al.~2005~\cite{2005MNRAS.359.1306W}, Read 2014~\cite{2014JPhG...41f3101R}, Silverwood et al.~2015~\cite{Silverwood:2015hxa}). 

\begin{figure}
    \begin{center}
        \includegraphics[width=\textwidth]{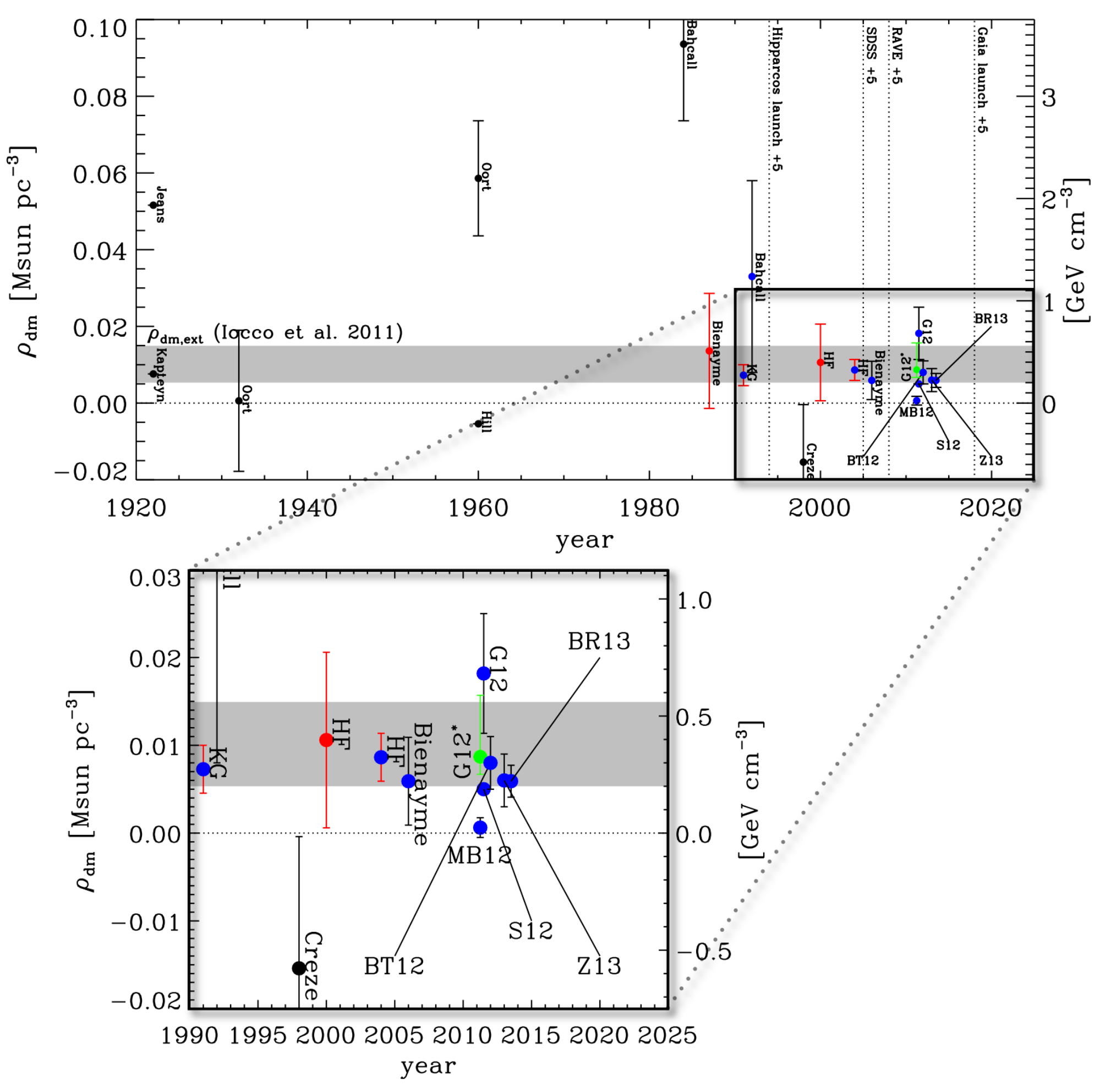}
    \caption{Timeline of local dark matter density measurements. See Read (2014) for further details and references~\cite{2014JPhG...41f3101R}.}
    \label{fig:justin}
    \end{center}
\end{figure}

\section{Dark Matter Particles}
\label{particles}

Over the past few decades, the very meaning of the phrase ``dark matter'' has evolved considerably. Today, this phrase is most frequently used as the name -- a proper noun -- of whatever particle species accounts of the bulk of our Universe's matter density. When a modern paper discusses the distribution of dark matter, or the impact of dark matter on structure formation, or the prospects for detecting dark matter with a gamma-ray telescope, the reader does not have to ask themselves whether the authors might have in mind white dwarfs, neutron stars, or cold clouds of gas -- they don't. This is in stark contrast to the earlier usage of the phrase, in which the word ``dark'' was a mere adjective, and ``dark matter'' included all varieties of astrophysical material that happened to be too faint to be detected with available telescopes. 

This linguistic transition reflects a larger change that has taken place over the past several decades within the astrophysics and particle physics communities. And although this transformation was driven and initiated by new scientific results and understanding, it also reflects a sociological change in the underlying scientific culture. Half a century ago, cosmology was something of a fringe-science, perceived by many astronomers and particle physicists alike to have little predictive power or testability. This can be easy to forget from our modern vantage point in the age of precision cosmology. Furthermore, prior to the last few decades, particle physicists did not often study or pursue research in astrophysics, and most astrophysicists learned and knew little about particle physics. As a result, these scientists did not frequently contribute to each other's fields of research. When Fermilab founded its theoretical astrophysics group in 1983, for example, the decision to do so was seen by many as a radical departure from the lab's particle physics mission. From the perspective of many particle physicists in the early 1980s, it was not obvious what astrophysics had to do with the questions being asked by particle physics. This view is shared by few today. As an illustration, we need only to note that the report of the US Particle Physics Project Prioritization Panel (P5) describes the ``Cosmic Frontier'', along with the Energy and Intensity Frontiers, as co-equal areas of inquiry within the larger field of particle physics. 

From our contemporary perspective, it can be easy to imagine that Fritz Zwicky, Vera Rubin, and the other early dark matter pioneers had halos of weakly interacting particles in mind when they discussed dark matter. In reality, however, they did not. But over time, an increasing number of particle physicists became interested in cosmology, and eventually in the problem of dark matter. By the late 1980s, the hypothesis that the missing mass consists of one or more yet-unknown subatomic particle species had gained enough support to become established as the leading paradigm for dark matter. As alternatives were ruled out one-by-one (see Chapters~\ref{baryonicdm} and~\ref{modgrav}), this view came to be held almost universally among both particle physicists and astrophysicists, as well as among their new and now increasingly common hybrids -- the particle-astrophysicists.  

\subsection{Neutrinos}
%\par
%\vspace{4\baselineskip}
%\centerline{-----------------------------------------------------------}
%%\centerline{*  *  *  *  *  *  *  *  *  *  *  *  *  *  *}
%\vspace{3\baselineskip}
%%\@afterindentfalse
%%\@afterheading

When one considers the dark matter problem from the perspective of the standard model of particle physics, the three neutrinos clearly stand out. Unlike all other known particle species, the neutrinos are stable -- or at least very long lived -- and do not experience electromagnetic or strong interactions. These are essential characteristics for almost any viable dark matter candidate. And although we know today that dark matter in the form of standard model neutrinos would be unable to account for our Universe's observed large scale structure, these particles provided an important template for the class of hypothetical species that would later be known as WIMPs -- weakly interacting massive particles. In this way, standard model neutrinos
%While the three known neutrino species may not have been the answer to the dark matter problem, they 
served as an important gateway particle, leading astrophysicists and particle physicists alike to begin their experimentation with a variety of other, more viable, particle dark matter candidates. And although the first scientists to consider the role of neutrinos in cosmology did not have the dark matter problem in mind -- many being unaware that there was any such problem to solve -- their work helped to establish the foundations that the field of particle dark matter would later be built upon. 

The earliest discussion of the role of neutrinos in cosmology appeared in a 1966 paper by S.~S.~Gershtein and Ya.~B.~Zeldovich~\cite{Gershtein:1966gg}. To many scientists working in fields of cosmology and particle-astrophysics, it will be no surprise to see Zeldovich's name attributed to this pioneering work. Yakov Borisovich Zeldovich was an utterly prolific and versatile physicist, making major contributions to the fields of material science, nuclear physics (including the Soviet weapons program), particle physics, relativity, astrophysics, and cosmology. In terms of research at the interface between particle physics and cosmology, it can sometimes seem like Zeldovich did almost everything first.

In the early 1960s, Zeldovich was one of only a handful of particle physicists who were also thinking about problems in cosmology. During this period, he made early contributions to black hole thermodynamics, recognized that accretion disks around black holes could power quasars, discussed the possibility of primordial black holes, and studied the problem of how the large scale structure of the Universe formed. He is probably most famous for his paper with Rashid Sunyaev, which predicted that the cosmic microwave background would be distorted by its inverse Compton scattering with high-energy electrons in galaxy clusters~\cite{Sunyaev:1970eu}. This so-called ``Sunyaev-Zeldovich effect'' was observed for the first time in 1983, and continues to be of considerable importance in modern cosmology.  So sweeping were Zeldovich's contributions to cosmology, that upon being introduced, Stephen Hawking is said to have expressed to him, ``Before I met you, I believed you to be a collective author, like Bourbaki\footnote{Nicolas Bourbaki was a collective pseudonym adopted by a group of 20th-century mathematicians.}.''

In their 1966 paper, Zeldovich and Gershtein considered the production of neutrinos under the conditions that existed shortly after the Big Bang. Making use of the knowledge of the newly discovered three degree cosmic microwave background~\cite{Penzias:1965wn}, they predicted how many electron and muon neutrinos would have existed in thermal equilibrium in the early Universe, and at what temperature those particles would have ceased to efficiently self-annihilate, leading to a population of neutrinos that survived as a thermal relic\footnote{As there existed no evidence for a third generation at the time, Gershtein and Zeldovich did not consider the tau neutrino.}. Considering how the density of those neutrinos would impact the expansion history of the Universe, and comparing that to existing estimates of the Hubble constant and the age of the oldest observed stars, Zeldovich and Gershtein concluded that the masses of the electron and muon neutrinos must each be less than approximately 400 eV; if they had been heavier, the neutrinos would have unacceptably slowed, or even reversed, the rate of cosmic expansion. For the muon neutrino, this result represented an improvement of three orders of magnitude over the previously existing upper limits.

Looking back at this result from a modern perspective, we see the seeds of particle dark matter, and even WIMPs. In particular, Zeldovich and Gershtein showed that a neutrino species with a mass of a few tens of eV or greater would come to dominate the energy density of the Universe. But there was no mention in their paper of any missing mass that these neutrinos might be able to account for; they only required that the density of the relic neutrinos not be so high as to cause the expansion rate of the Universe to slow down faster than observed. 

This is essentially the same perspective that was expressed years later, when papers on this topic began to appear in the West. The first of these papers appeared in 1972, in which Ram Cowsik and J.~McClelland used an approach similar to Zeldovich and Gershtein's to derive an upper limit of 8 eV on the mass of a single (Dirac) neutrino species~\cite{Cowsik:1972gh} (see also Ref.~\cite{neutrino72}). If it had not been for this paper, one might be tempted to conclude that interest in this topic would have developed much sooner among American and Western European scientists if word of Zeldovich and Gershtein's work had reached them earlier.
%
%if only word of Zeldovich and Gershtein's work had been better disceminated to their European or American colleagues, interest in this topic would have developed much sooner in those parts of the world. 
But the paper by Cowsik and McClelland (who were both at the University of California, Berkeley, at the time) seems to disprove this counterfactual. Even after the appearance of this paper, there was no discernible rush to further explore the role of neutrinos (or other thermal relics) in the early Universe.

Eventually, however, interest in neutrino cosmology did begin to pick up. In 1976, A.~S.~Szalay and G.~Marx published a paper that not only derived an upper limit on neutrino masses from cosmology, but also discussed the possibility that $\sim$10 eV neutrinos might make up the ``missing mass'' in the Universe, and in galaxy clusters. Then, a few years later, a sequence of related papers appeared in rapid succession. In a paper received in April of 1977, Piet Hut presented a limit on the neutrino mass from cosmological considerations, ruling out masses in the range of 120 eV to 3 GeV~\cite{Hut:1977zn}. In contrast to the authors of the preceding papers, Hut pointed out that quite heavy neutrinos ($m_{\nu}>$~3 GeV) would be produced in the Big Bang with an abundance that would not overclose the Universe. Only about a week later, Ben Lee and Steven Weinberg submitted a paper that included a very similar lower bound ($m_{\nu}>2$ GeV)~\cite{Lee:1977ua}. In the same month, a paper by K.~Sato and H.~Kobayashi~\cite{Sato:1977ye} presented similar conclusions, and another by Duane Dicus, Edward ``Rocky'' Kolb and Vigdor Teplitz pointed out that such bounds could be evaded if neutrinos were unstable~\cite{Dicus:1977nn}. A month later, a new paper by Zeldovich (with M.~I.~Vysotskii and A.~D.~Dolgov) appeared, updating their own cosmological constraints on neutrino mass~\cite{Vysotsky:1977pe}. 

Despite the very interesting and important results of these papers, it is notable that most of them did not attempt to address, or even acknowledge, the possibility that neutrinos could account for the missing mass observed by astronomers on galactic and cluster scales. Exceptions to this include the 1976 paper of Szalay and Marx, and the 1977 paper of Lee and Weinberg, whose final sentence reads as follows~\cite{Lee:1977ua}: 
\begin{quote}
\emph{Of course, if a stable heavy neutral lepton were discovered with a mass of order 1-15 GeV, the gravitational field of these heavy neutrinos would provide a plausible mechanism for closing the universe.}
\end{quote}

While this is still a long way from acknowledging the dynamical evidence for dark matter, it was an indication that physicists were beginning to realize that weakly interacting particles could be very abundant in our Universe, and may have had an observable impact on its evolution. The connection between particle physics and the missing mass problem did gradually become more appreciated over the years to come. In 1978, for example, a paper by James Gunn, Ben Lee, Ian Lerche, David Schramm, and Gary Steigman included the following statement in their abstract~\cite{Gunn:1978gr}:
\begin{quote}
\emph{...\,such a lepton is an excellent candidate for the material in galactic halos and for the mass required to bind the great clusters of galaxies.}
\end{quote}

By the end of the decade, a number of scientists -- including Zeldovich and his Moscow group~\cite{Doroshkevich:1980zs,Doroshkevich:1980vy,Zeldovich:1980st} -- had begun to argue in favor of neutrinos as dark matter. Interest in this possibility grew considerably in and after 1980, when a group studying tritium beta decay reported that they had measured the mass of the electron anti-neutrino (and presumably also the electron neutrino) to be approximately 30 eV~\cite{30eV}. With a mass of this value, neutrinos would be expected to have played a very significant role in cosmology. And although this ``discovery'' was eventually refuted, it motivated many particle physicists to further investigate the cosmological implications of their research, and encouraged many astrophysicists to consider the possibility that the dark matter halos surrounding galaxies and galaxy clusters might not be made up of faint stars or other astrophysical objects, but instead might consist of a gas of non-baryonic particles. 

By the middle of the 1980s, a new tool had come into use that would put neutrino dark matter to the test. This tool --- numerical simulations --- could be used to predict how large numbers of dark matter particles would evolve under the force of gravity in an expanding Universe, and thus was able to assess the cosmological role and impact of dark matter particles on the formation of large scale structure. Importantly, such tests could be used to discriminate between different dark matter candidates, at least in some cases. 

%By comparing the results of these simulations with the observed distributions of galaxies and galaxy clusters, it became possible to begin to narrow down the list of candidates for the particle species that makes up the Universe's dark matter.

The primary characteristic of a given particle dark matter candidate that can be probed by numerical simulations is whether it was relativistic (hot) or non-relativistic (cold) during the epoch of structure formation\footnote{The terms ``hot'' and ``cold'' dark matter were coined in 1983 by Joel Primack and Dick Bond (J.~Primack, private communication).}. Standard model neutrinos, being very light thermal relics, are predicted to emerge from the early Universe with a highly relativistic velocity distribution, and thus represent an example of hot dark matter~\cite{1981ApJ...243....1S,1982ApJ...258..415P}. Simulations have shown that hot dark matter particles would tend to collapse and form very large structures first, and only later go on to form smaller (i.e.~galaxy-sized) halos through the fragmentation of larger halos. In contrast to this ``top-down'' sequence of structure formation, cold dark matter particles form structures through a ``bottom-up'' sequence, beginning with the smallest halos, which go on to form larger halos through a succession of mergers. 

From these early simulations, it quickly became clear that hot and cold dark matter lead to very different patterns of large scale structure. By comparing the results of these simulations with those of galaxy surveys (in particular the CfA survey, which was the first extensive 3D survey of galaxies in the local Universe~\cite{cfa}), it was determined that standard model neutrinos -- or any other examples of hot dark matter -- could not account for most of the dark matter in the Universe~\cite{hdmfails}. In their 1983 paper, Simon White, Carlos Frenk and Marc Davis make the following statement about a neutrino-dominated Universe~\cite{hdmfails}:
\begin{quote}
\emph{We find [the coherence length] to be too large to be consistent with the observed clustering scale of galaxies... The conventional neutrino-dominated picture appears to be ruled out.}
\end{quote}

We will discuss numerical simulations, and their role in the history of dark matter, in greater detail in Sec.~\ref{sec:simulations}.

As it became accepted that standard model neutrinos could not make up most of the Universe's dark matter\footnote{A possible exception being the tau neutrino, whose mass would not be measured for another two decades, and thus could not at the time be ruled out as a cold dark matter candidate.}, it also became clear that there must exist at least one currently unknown particle species that makes up the missing mass. But although standard model neutrinos were far too light and hot to make up the dark matter, this new information did not preclude the possibility that other types of neutrino-like particles might make up this elusive substance (see, for example, Ref.~\cite{Olive:1981ak}). In 1993, Scott Dodelson and Lawrence Widrow proposed a simple scenario in which an additional neutrino species, without the electroweak interactions experienced by standard model neutrinos, could be produced in the early Universe and realistically make up the dark matter~\cite{Dodelson:1993je}. Other than through gravity, the particles envisioned by Dodelson and Widrow interact only through a small degree of mixing with the standard model neutrinos. With such feeble interactions, such particles would have never been in thermal equilibrium in the early Universe, but instead would have been produced through the oscillations of the other neutrino species. Depending on their mass, such sterile neutrinos could be produced with a wide range of temperatures, and thus could constitute either a warm ($m_{\nu_s} \sim$ keV) or a cold ($m_{\nu_s} \gg$ keV) candidate for dark matter.

\subsection{Supersymmetry}

Among the particle species contained within the standard model, neutrinos are the only examples that are stable, electrically neutral, and not strongly interacting, and therefore are the only known particles that were viewed as potentially viable candidates for dark matter. Physicists' imagination, however, would not remain confined to the standard model for long, but instead would turn to the contemplation of many speculative and yet undiscovered candidates for the dark matter of our Universe. In particular, beginning in the early 1970s, many physicists began to consider the possibility that nature may contain a spacetime symmetry relating fermions to bosons, dubbed ``supersymmetry''~\cite{Gervais:1971ji,Golfand:1971iw,earlysusy3,Wess:1974tw}. Supersymmetry requires that for every fermion, a boson must exist with the same quantum numbers, and vice versa. Supersymmetry, therefore, predicts the existence of several new electrically neutral and non-strongly interacting particles, including the superpartners of the neutrinos, photon, $Z$ boson, Higgs boson, and graviton. If any of these superpartners were stable, they could be cosmologically abundant, and may have played an important role in the history and evolution of our Universe. 

The cosmological implications of supersymmetry began to be discussed as early as the late 1970s. In Piet Hut's 1977 paper on the cosmological constraints on the masses of neutrinos (as described above), the discussion was not entirely limited to neutrinos, or even to weakly interacting particles. Even the abstract of that paper mentions another possibility~\cite{Hut:1977zn}:
\begin{quote}
\emph{Similar, but much more severe, restrictions follow for particles that interact only gravitationally. This seems of importance with respect to supersymmetric theories.}
\end{quote}

The paper goes on to close with the first cosmological bounds on the mass of the supersymmetric partner of the graviton, the spin $3/2$ gravitino:
\begin{quote}
\emph{Assuming the standard big bang model to be relevant in the context of supergravity theories, one can make the following remark. If there exist light massive spin $3/2$ particles interacting only gravitationally, having four spin degrees of freedom, their mass must be less than 15 eV if they are their own antiparticles, otherwise their mass is less than 1.5 eV. Also, they may exist with masses very much larger than 1 TeV.}
\end{quote}
Although such bounds would be revised in the decades to follow, in particular being shown to depend on the temperature to which the Universe was reheated following inflation, this result is essentially the basis of what is known today as the ``cosmological gravitino problem''. 

In their 1982 paper, Heinz Pagels and Joel Primack also considered the cosmological implications of gravitinos~\cite{Pagels:1981ke}. But unlike Hut's paper, or the other preceding papers that had discussed neutrinos as a cosmological relic, Pagels and Primack were clearly aware of the dark matter problem, and explicitly proposed that gravitinos could provide the solution by making up the missing mass~\cite{Pagels:1981ke}:
\begin{quote}
\emph{Gravitinos could also provide the dark matter required in galactic halos and small clusters of galaxies.}
\end{quote}

In many ways, Pagel and Primack's letter reads like a modern paper on supersymmetric dark matter, motivating supersymmetry by its various theoretical successes and attractive features, and going on to discuss not only the missing mass in galaxies and clusters, but also the role that dark matter could play in the formation of large scale structure. At the time of Pagel and Primack's submission, however, supersymmetry itself had not yet taken its modern form, and no truly realistic supersymmetric models had been proposed (although many important steps had been made in this direction~\cite{Fayet:1974pd,Fayet:1976et,Fayet:1977vd,Fayet:1977yc}). This changed in December of 1981, when a paper by Savas Dimopoulos and Howard Georgi described a model that would become known as the minimal supersymmetric standard model (MSSM)~\cite{Dimopoulos:1981zb}. 

The advent of the MSSM opened the door to considering superpartners other than the gravitino as cosmological relics. In particular, in the MSSM, the superpartners of the photon, the $Z$, and two neutral scalar Higgs bosons mix to form four particles that would become known as neutralinos. Over the past three and a half decades, neutralinos have been the single most studied candidate for dark matter, having been discussed in many thousands of scientific publications. In order to be the dark matter, however, something must stabilize the lightest neutralino, preventing these particles from decaying shortly after being created. 

In supersymmetric extensions of the standard model, there exist interactions that violate the conservation of baryon and lepton number. Unless the relevant couplings are highly suppressed, such interactions are expected to cause the proton to decay on unacceptably short timescales, on the order of a year or less. It was recognized early in supersymmetry's development, however, that the proton's lifetime could be made to safely exceed observational limits if an additional -- and well-motivated -- symmetry known as $R$-parity~\cite{Fayet:1974pd,Salam:1974xa,Fayet:1977yc,Farrar:1978xj} is imposed. The $R$-parity of a given particle is defined as follows:
\begin{equation}
P_R =  (-1)^{2s+3B+L},
\end{equation}
where $s$ is the spin of the particle, and $B$ and $L$ are the particle's baryon number and lepton number, respectively. Under this definition, all of the standard model particles have positive $R$-parity, $P_R=+1$, while all of their superpartners have $P_R=-1$. As a consequence, this parity ensures that superpartners can only be created or destroyed in pairs. A heavy superpartner can decay into a lighter superpartner, along with any number of standard model particles, but the lightest of the superpartners cannot decay. Thus if the lightest superpartner of the MSSM is either a neutralino or a sneutrino (the superpartner of a standard model neutrino), $R$-parity will stabilize it, allowing it to be a potentially viable dark matter candidate. As far as we are aware, it was Pagels and Primack who were the first to invoke $R$-parity in order to stabilize a dark matter candidate~\cite{Pagels:1981ke}.

Papers discussing the cosmological implications of stable neutralinos began to appear in 1983\footnote{Unstable but long-lived photinos had been considered earlier, in 1981, by Nicola Cabibbo, Glennys Farrar and Luciano Maiani~\cite{Cabibbo:1981er}}. In the first two of these papers, Steven Weinberg~\cite{Weinberg:1982tp} and Haim Goldberg~\cite{Goldberg:1983nd} independently discussed the case of a photino -- a neutralino whose composition is dominated by the superparter of the photon -- and derived a lower bound of 1.8 GeV on its mass by requiring that the density of such particles does not overclose the Universe. A few months later, a longer paper by John Ellis, John Hagelin, Dimitri Nanopoulos, Keith Olive and Mark Srednicki considered a wider range of neutralinos as cosmological relics~\cite{Ellis:1983ew}. In Goldberg's paper, there is no mention of the phrase dark matter or of any missing mass problem, and Ellis et al.~took a largely similar approach, simply requiring that the cosmological abundance of neutralinos not be so large as to overly slow or reverse the Universe's expansion rate. Ellis et al., however, did mention the possibility that neutralinos could make up the dark matter, although only in a single sentence~\cite{Ellis:1983ew}:
\begin{quote}
\emph{A more restrictive constraint follows from the plausible assumption that a non-relativistic [supersymmetric] fermion would participate in galaxy formation, in which case the limits on ``dark matter'' in galaxies allow one to deduce that $\rho_{\chi}\leq 2\times 10^{-30} \, (\Omega h^2)$ gm/cm$^3$.}
\end{quote}
and in a passing footnote of Ref.~\cite{Ellis:1983wd}:
\begin{quote}
\emph{This bound comes from the overall density of the universe and is very conservative. One can argue that massive neutral fermions probably condense into galaxies in which case a more stringent limit coming from missing galactic matter could be applied.}
\end{quote}

Although far from a full embrace of a particle physics solution to the dark matter problem, these sentences (along with those expressed by Pagels and Primack~\cite{Pagels:1981ke}, and by Jim Peebles within the context of massive neutrinos~\cite{1982ApJ...258..415P}) reflected the emergence of a new perspective\footnote{Early evidence for this transition can found in the conferences that took place over this period of time, including the ``Study Week on Cosmology and Fundamental Physics'', that was held at the Vatican in September and October of 1981.}. Throughout the decades to follow, a countless number of particle physicists would motivate their proposals for physics beyond the standard model by showing that their theories could account for the Universe's dark matter. Despite any other attractive features that a given theory might possess, if it cannot provide a dark matter candidate, it would come to be viewed as incomplete.

That supersymmetric particles, and the lightest neutralino in particular, have received so much attention as dark matter candidates is due, in large part, to the fact that the motivation for supersymmetry does not primarily rely on the dark matter problem. Particle physicists have been drawn to supersymmetry over the past four decades for its ability to solve the electroweak hierarchy problem, and to enable gauge coupling unification~\cite{Dimopoulos:1981yj,Ibanez:1981yh,Marciano:1981un}, combined with its unique nature as both a spacetime symmetry and an internal symmetry~\cite{Haag:1974qh}. If in some other universe, astrophysicists had measured the cosmological density of matter to be consistent with the observed density of stars, gas, and other baryons, particle physicists in that universe may have been just as interested in supersymmetry as they are in ours. In this respect, supersymmetry's ability to provide a viable dark matter candidate is seen by many particle physicists as something of a bonus, rather than as the primary motivation to study such theories.

Supersymmetry, however, is not the only particle physics framework that is both strongly motivated in its own right, and able to provide a viable candidate for the dark matter of our Universe. In the next section, we will turn our attention to perhaps the second most studied candidate for dark matter, the axion.

\subsection{Axions}

By all measures, quantum chromodynamics (QCD) has been an incredibly successful theory, and describes the strong force and the quarks and gluons which experience it with remarkable precision. That being said, QCD does suffer from one troubling issue, known as the strong-CP problem.  This problem comes down to the fact that the QCD Lagrangian contains the following term:
\begin{equation}
{\mathcal L}_{\rm QCD} \supset \bar{\Theta}\, \frac{g^2}{32 \pi^2} \, G^{a \mu \nu} \tilde{G}_{a \mu \nu},
\end{equation}
where $G^{a \mu \nu}$ is the gluon field strength tensor and $\bar{\Theta}$ is a quantity closely related to the phase of the QCD vacuum. If $\bar{\Theta}$ were of order unity, as would naively be expected, this term would introduce large charge-parity (CP) violating effects, causing the electric dipole moment of the neutron to be $\sim$$10^{10}$ times larger than experimental upper bounds permit. Therefore, to be consistent with observations, the quantity $\bar{\Theta}$ must be smaller than $\sim$$10^{-10}$. While this could be nothing more than a highly unlikely coincidence, it has been interpreted by many as an indication that some new physics comes in to explain why $\bar{\Theta}$ is so small. This is the essence of the strong-CP problem.

What is perhaps the most promising solution to this problem was proposed in 1977 by Roberto Peccei and Helen Quinn~\cite{Peccei:1977ur,Peccei:1977hh}. They showed that by introducing a new global $U(1)$ symmetry that is spontaneously broken, the quantity $\bar{\Theta}$ can be dynamically driven toward zero, naturally explaining the small observed value. Later in the same year, Frank Wilczek~\cite{Wilczek:1977pj} and Steven Weinberg~\cite{Weinberg:1977ma} each independently pointed out that such a broken global symmetry also implies the existence of a Nambu-Goldstone boson, called the axion. The axion acquires a small mass as a result of the $U(1)$ symmetry's chiral anomaly, on the order of $m_a \sim \lambda^2_{\rm QCD}/f_{\rm PQ}$, where $f_{\rm PQ}$ is the scale at which the symmetry is broken. 

In its original conception, $f_{\rm PQ}$ was taken to be near the weak scale, leading to an MeV-scale axion mass. This scenario was quickly ruled out, however, by a combination of laboratory and astrophysical constraints. In particular, in contradiction with observation, axions heavier than $\sim$10 keV are predicted to induce sizable rates for a number of exotic meson decays, such as $K^+ \rightarrow \pi^+ + a$ and $J/\psi \rightarrow \gamma + a$. Similarly, axions heavier than $\sim$1 eV would lead to the very rapid cooling of red giant stars, again in contradiction with observations. Some years later, after the occurrence and observation of Supernova 1987A, even stronger constraints were placed on the axion mass, $m_a \lsim 10^{-3}$ eV.  

In order to evade these constraints, axions must be much lighter, and much more feebly interacting~\cite{1979PhRvL..43..103K,Shifman:1979if,Dine:1981rt}, than had been originally envisioned by Wilczek and Weinberg. Such light and ``invisible'' axions, however, can have very interesting consequences for cosmology. Being stable over cosmological timescales, any such axions produced in the early Universe will survive and, if sufficiently plentiful, could constitute the dark matter. 

A number of mechanisms have been considered for the production of axions in the early Universe. As with other particle species, axions can be produced thermally~\cite{Turner:1986tb,Kephart:1986vc}. For axions light enough to avoid the above mentioned constraints, however, the thermal relic abundance is predicted to be very small, and would only be able to account for a small fraction of the dark matter density. There is, however, another production mechanism, related to the misalignment of the Peccei-Quinn field, that is likely to be more important in the mass range of interest~\cite{Dine:1982ah,Abbott:1982af,Preskill:1982cy}. Although the quantity $\bar{\Theta}$ is dynamically driven to zero by the mechanism proposed by Peccei and Quinn, its initial value was likely to be some much larger value, presumably determined through some random process. As the temperature of the Universe dropped below $T \sim \lambda_{\rm QCD}$, and the value of $\bar{\Theta}$ was driven toward zero, the energy that had been stored in the Peccei-Quinn field gets transferred into the production of a non-thermal axion population. For typical initial conditions, this process of misalignment production is predicted to generate a density of axions that is comparable to the dark matter density for masses on the order of $m_a \sim 10^{-5}$ eV. Alternatively, it was pointed out that as a consequence of $\bar{\Theta}$ taking on different initial values in different locations throughout space, a network of topological defects (axionic strings and domain walls) may be expected to form. The subsequent decay of these defects is predicted to generate a quantity of axions that is comparable to that resulting from misalignment production~\cite{Davis:1986xc}. Inflation will erase this network of topological defects, however, unless it occurs prior to the breaking of the Peccei-Quinn symmetry.

In light of these considerations, axions with masses in the range of $m_a \sim 10^{-6}-10^{-4}$ eV, and generated largely via misalignment production, have become one of the most popular and well-studied candidates for dark matter. Alternatively, it was also pointed out that if inflation occurs after the breaking of the Peccei-Quinn symmetry, then there may also be a viable anthropic scenario in which the axion mass could be much lighter~\cite{Linde:1991km,Wilczek:2004cr,Tegmark:2005dy}. In this scenario, the initial value of $\bar{\Theta}$ is of order unity in most regions, leading to very high axion densities and to the rapid contraction of space. 
%
%within the regions of space in which $\bar{\Theta}$ initially takes on typical (of order unity) values, the cosmological density of axions would be very high, strongly overclosing the Universe. 
%
In a small fraction of the overall cosmic volume, however, the initial value of $\bar{\Theta}$ will be much lower, leading to far less axion production. If we speculate that life is only able to emerge in those regions in which the Universe is allowed to expand for millions or billions of years or more, we should expect to find ourselves in a region with a density of axions that is similar to the observed density of dark matter, even if the axion is much lighter than non-anthropic estimates would lead us to expect.

\subsection{The WIMP Paradigm}

By the end of the 1980s, the conclusion that most of the mass in the Universe consists of cold and non-baryonic particles had become widely accepted, among many astrophysicists and particle physicists alike. And while alternatives continued to be discussed (see the following two chapters), cold dark matter in the form of some unknown species of elementary particle had become the leading paradigm. In addition to massive neutrinos (sterile or otherwise), supersymmetric particles (neutralinos, gravitinos, sneutrinos, axinos) and axions were each widely discussed as prospective dark matter candidates. And as the evidence in favor of non-baryonic dark matter became increasingly compelling, an ever greater number of particle physicists began to openly speculate about the nature of this invisible substance. The result of this was a long and diverse list of exotic possibilities, ranging from topological defects produced through spontaneous symmetry breaking in the early Universe (monopoles, cosmic strings)~\cite{Kibble:1976sj}, to macroscopic configurations of quark matter (centimeter-scale ``nuggets'', with nuclear-scale densities)~\cite{Witten:1984rs}, and even ``pyrgons'' (Kaluza-Klein excitations) that could appear within the context of models with extra spatial dimensions~\cite{Kolb:1983fm}. 

While this proliferation of dark matter candidates was taking place, however, a commonality among many of the proposed particles was becoming increasingly appreciated. In order for a particle species to freeze-out of thermal equilibrium in the early Universe to become a cold relic, it must not be too light (roughly heavier than $\sim$1-100 keV). Furthermore, for the predicted thermal relic abundance of such a species to match the observed dark matter density, the dark matter particles must self-annihilate with a cross section on the order of $\sigma v \sim 10^{-26}$ cm$^3$/s (where $v$ is the relative velocity between the annihilating particles). This number is strikingly similar to the cross section that arises from the weak force. For example, a stable neutrino with a mass of several GeV, annihilating through the exchange of a $Z$-boson, would freeze-out with a relic abundance that is roughly equal to the measured density of dark matter. Furthermore, such conclusions are not limited to neutrinos, but apply to a broad range of electroweak-scale dark matter candidates -- including any number of stable particles with MeV-TeV masses and interactions that are mediated by the exchange of electroweak-scale particles.
%
%A generic stable particle with an MeV-TeV mass (neutralinos, sneutrinos, massive neutrinos, etc.) and that interacts through the exchange of particles with a mass on the order of the electroweak scale (of the order of $10^2$ GeV) will act as a cold relic with roughly the observed dark matter abundance. 
%
This observation, combined with theoretical arguments in favor of the existence of new physics at or around the electroweak scale, have elevated weakly interacting massive particles (WIMPs)~\cite{Steigman:1984ac} to the leading class of candidates for dark matter\footnote{Although the term WIMP, as coined by Gary Steigman and Michael Turner in 1984, was originally intended to include all particle dark matter candidates, including axions, gravitinos, etc., the definition of this term has since evolved to more often denote only those particles that interact through the weak force.}. WIMPs have been the subject of thousands of theoretical studies, leading to the refinement of many calculations, including that of the dark matter's thermal relic abundance~\cite{Srednicki:1988ce,Gondolo:1990dk,Griest:1990kh}. Furthermore, WIMPs (and to a somewhat lesser degree, axions) have motivated an expansive experimental program that continues to this day. With the advent of the Large Hadron Collider at CERN, and ever more sensitive astrophysical experiments, many believe that the moment of truth has come for WIMPs: either we will discover them soon, or we will begin to witness the decline of the WIMP paradigm~\cite{Bertone:2010at}.

% CONVENTIONS:
% standard model
% the/our Universe

\section{Baryonic Dark Matter}
\label{baryonicdm}

As the evidence in favor of dark matter in galaxies and galaxy clusters accumulated, more and more astronomers began to contemplate what might make up this faint material. To many astronomers and astrophysicists, the most obvious possibility was that this missing mass might consist of compact objects that were much less luminous than -- but otherwise qualitatively similar to -- ordinary stars. Possibilities for such objects included planets, brown dwarfs, red dwarfs, white dwarfs, neutron stars, and black holes. Kim Griest would later coin the term ``MACHOs'' -- short for massive astrophysical compact halo objects -- to denote this class of dark matter candidates, in response to the leading alternative of weakly interacting massive particles, ``WIMPs''.

Although there is a consensus today that MACHOs do not constitute a large fraction of the dark matter, opinions differ as to which lines of evidence played the most important role in reaching that conclusion (for an example of some of the very early arguments that had been made against MACHOs as dark matter, see Ref.~\cite{1983PhLB..126...28H}). That being said, two lines of investigation would ultimately prove to be particularly important in resolving this question: searches for MACHOs using gravitational microlensing surveys, and determinations of the cosmic baryon density based on measurements of the primordial light element abundances and of the cosmic microwave background.

% The first of these were the results of microlensing surveys, which failed to detect the number of MACHOs required to account for the Milky Way's dark halo. Over a similar period of time, cosmological observations (light element abundances and the cosmic microwave background) began to exclude the possiblity that the universe could contain enough baryons to make up much of the dark matter. 

\subsection{Gravitational Microlensing}

The possibility that light could be deflected by gravity has a long history, extending back as far as Newton. In 1915, Einstein made the correct prediction for this phenomena using the framework of general relativity (which predicts twice the degree of deflection as Newtonian gravity). An early test of general relativity was famously conducted during the solar eclipse of 1919, which provided an opportunity to measure the bending of light around the Sun. Although the measurements obtained by Arthur Eddington favored the relativistic prediction, other simultaneous observations appeared to agree with the Newtonian expectation. Despite this apparent ambiguity, Eddington's results were seen as persuasive by many astronomers, and served to elevate the status of Einstein's theory.

In 1924, the Russian physicist Orest Chwolson returned to the topic of gravitational lensing, pointing out that a massive body could deflect the light from a more distant source in such a way that would lead to the appearance of multiple images, or of a ring~\cite{1924AN....221..329C}. In 1936, Einstein himself published a paper on this topic~\cite{1936Sci....84..506E}, but concluded that due to the very precise alignment required, ``there is no great chance of observing this phenomenon''.

The modern theory of gravitational lensing was developed in the 1960s, with contributions from Yu Klimov~\cite{1963AZh....40..874K,1963SPhD....8..119K,1963SPhD....8..431K}, Sidney Liebes~\cite{1964PhRv..133..835L}, and Sjur Refsdal~\cite{1964MNRAS.128..295R,1964MNRAS.128..307R}, followed by the first observation of a lensed quasar by Dennis Walsh, Robert Carswell and Ray Weyman in 1979~\cite{1979Natur.279..381W}. In the same year, Kyongae Chang and Sjur Refsdal showed that individual stars could also act as lenses, leading to potentially observable variations over timescales of months~\cite{1979Natur.282..561C}. In 1986, Bohdan Paczynski proposed that this phenomena of gravitational microlensing could be used to search for compact objects in the ``dark halo'' of the Milky Way~\cite{1986ApJ...304....1P}, followed in 1987 by more detailed predictions for the probability and light curves of such events, described in the Ph.D. thesis of Robert Nemiroff~\cite{1987PhDT........12N}\footnote{The possibility that objects in the Milky Way's dark halo could be detected through gravitational lensing was also discussed earlier, in a chapter of the 1981 Ph.D. thesis of Maria Petrou. On the advice of her supervisor, Petrou did not otherwise attempt to publish this work~\cite{VallsGabaud:2012xz}.}. 

The strategy proposed by these authors was to simultaneously monitor large numbers of stars in a nearby galaxy (such as in the Large Magellanic Cloud), in an effort to detect variations in their brightness. If the halo consisted entirely of MACHOs, approximately one out of 2 million stars should be magnified at a given time, a ratio known as the microlensing optical depth. Furthermore, as the duration of a microlensing event is predicted to be $t \sim 130 \, {\rm days} \times (M/M_{\odot})^{0.5}$, such a program would be best suited to detect objects with masses in the range of $\sim$$10^{-7} M_{\odot}$ to $\sim$$10^2 M_{\odot}$, corresponding to variations over timescales of hours to a year. These factors motivated the approaches taken by the MACHO, EROS (Experience pour la Recherche d'Objets Sombres), and OGLE (Optical Gravitational Lensing Experiment) Collaborations, who each set out to conduct large microlensing surveys in order to test the hypothesis that the Milky Way's dark halo consisted of MACHOs. 

Although the first claim of a microlensing event was reported in 1989, by Mike Irwin and collaborators~\cite{1989AJ.....98.1989I}, the implications of microlensing surveys for dark matter only began to take shape a few years later with the first results of the MACHO Collaboration. The MACHO Collaboration was a group of mostly American astronomers making use of the 1.27-meter telescope at the  Mount Stromlo Observatory in Australia to simultaneously monitor millions of stars in the Large Magenellic Cloud. In October of 1993, they reported the detection of their first microlensing event, consistent with a 0.03 to 0.5 $M_{\odot}$ MACHO~\cite{1993Natur.365..621A}. In the same month, the EROS Collaboration reported the detection of two such events, favoring a similar range of masses~\cite{1993Natur.365..623A}. At the time, the rate of these events appeared to be consistent with that anticipated from a halo that was dominated by MACHOs. Kim Griest (a member of the MACHO Collaboration) recalled in 2000:
\begin{quote}
\emph{After the discovery of MACHOs in 1993, some thought that the dark matter puzzle had been solved.}
\end{quote}
But alas, it was not to be.

Over a period of 5.7 years, the MACHO Collaboration measured the light curves of 40 million individual stars, identifying between 14 and 17 candidate microlensing events. This was well above their expected background rate, and lead them to conclude that between 8\% and 50\% of the Milky Way's halo mass consisted of compact objects, most of which had masses in the range of 0.15 to 0.9 $M_{\odot}$~\cite{2000ApJ...542..281A}. After collecting data for 6.7 years, however, the EROS Collaboration had identified only one microlensing candidate event, allowing them to place an upper limit of 8\% on the halo mass fraction in MACHOs~\cite{2000A&A...355L..39L,2007A&A...469..387T}. Compact objects, at least within the mass range probed by microlensing surveys, do not appear to dominate the missing mass in the Milky Way's halo.

\subsection{The Universe's Baryon Budget}

Throughout much of the mid-twentieth century, the origin of the various nuclear species remained a subject of considerable mystery and speculation. As early as 1920, Arthur Eddington and others argued that the fusion of hydrogen into helium nuclei could be capable of providing the primary source of energy in stars, and suggested that it might also be possible to generate heavier elements in stellar interiors~\cite{1920Obs....43..341E,1920Natur.106...14E}. In 1939, Hans Bethe expanded significantly upon this idea, describing the processes of the proton-proton chain and the carbon-nitrogen-oxygen cycle that are now understood to dominate the energy production in main sequence stars~\cite{1939PhRv...55..434B}. Fred Hoyle, in papers in 1946 and 1954, calculated that nuclei as heavy as iron could be synthesized in massive stars~\cite{1946MNRAS.106..343H}, and that even heavier nuclear species could be produced by supernovae~\cite{1954ApJS....1..121H}. 
 
An alternative to stellar nucleosynthesis was proposed in 1946 by George Gamow~\cite{1946PhRv...70..572G}, and followed up upon two years later in a paper by Gamow and Hermann Alpher~\cite{1948PhRv...73..803A}. The author list of this later paper also famously included Hans Bethe (who reportedly did not contribute to the research) in order to facilitate the pun that enabled it to become known as the ``alpha-beta-gamma'' paper. In this pair of papers, it was proposed that all nuclear species (both light and heavy) may have been produced in the early Universe through the process of neutron capture. While of historic significance, there were considerable technical problems with the calculations presented in these early papers, some of which were pointed out by Enrico Fermi, Chushiro Hayaski, and Anthony Turkevich in the years to follow. Among other flaws, Alpher and Gamow did not correctly account for Coulomb barriers in estimating the rates for nuclear fusion. Perhaps more importantly, they did not appreciate that the lack of stable nuclei with atomic numbers in the range of 5-8 would effectively prevent any significant nucleosynthesis from occurring beyond $^4$He. After accounting for these issues, Alpher, along with Robert Herman and James Follin, correctly predicted the abundance of helium produced in the early Universe, and reported in 1953 that the heavier elements could not be accounted for by this mechanism~\cite{1953PhRv...92.1347A}. For these and other reasons, stellar nucleosynthesis remained the predominant theory throughout the 1950s and into the 1960s. That being said, by the late 1950s, it was becoming increasingly clear that stellar nucleosynthesis could not generate enough helium to accommodate the observed abundance, as summarized in the classic 1957 review paper by Margaret Burbidge, Geoffrey Burbidge, William Fowler, and Fred Hoyle~\cite{1957RvMP...29..547B}.

The discovery of the cosmic microwave background in 1965 lead to increased interest in Big Bang nucleosynthesis, and made it possible to further refine the predictions for the light element abundances. In particular, the temperature of this newly detected background favored a primordial helium fraction in the range of 26-28\%~\cite{1966ApJ...146..542P,1967ApJ...148....3W}, consistent with observations. In 1973, a paper by Hubert Reeves, Jean Audouze, William Fowler and David Schramm focused on the production of deuterium in the early Universe~\cite{1973ApJ...179..909R}. As deuterium had been detected in the interstellar medium, but is not generated in stars, these authors argued that Big Bang nucleosynthesis offered the most plausible origin for the observed deuterium. In the same paper, the authors also used the measured light element abundances to derive an upper limit on the cosmological baryon density that was about one tenth of the critical density, $\Omega_b \lsim 0.1 \, \Omega_{\rm crit}$. 

Constraints on the cosmological baryon density became increasingly stringent over the decades to follow. Of particular importance were the first high-precision measurements of the primordial deuterium abundance, which were carried out in the late 1990s by Scott Burles, David Tytler, and others~\cite{1998ApJ...507..732B,1998ApJ...499..699B,2001ApJ...552..718O}. These measurements were used to determine the baryonic abundance with roughly 10\% precision, $\Omega_b h^2 = 0.020 \pm 0.002$ (95\% CL)~\cite{2001ApJ...552L...1B}; leaving little room for baryonic MACHOs~\cite{Fukugita:1997bi}. At around the same time, measurements of the angular power spectrum of the cosmic microwave background were also becoming sensitive to this quantity. In particular, the ratio of the heights of the odd and even peaks in this power spectrum is primarily set by the baryonic density. Although limited measurements of the second peak were made by ground- and balloon-based experiments in the late 1990s, it was not until the satellite-based WMAP experiment that these determinations became competitive with (and superior to) those based on the measured light element abundances. WMAP ultimately achieved a measurement of  $\Omega_b h^2 = 0.02264 \pm 0.00050$ (68\% CL)~\cite{2013ApJS..208...19H}, while the most recent analysis from the Planck Collaboration arrives at a constraint of $\Omega_b h^2 = 0.02225 \pm 0.00016$, corresponding to a fractional uncertainty of less than one percent~\cite{2015arXiv150201589P}. When this is compared to the total matter density as inferred by these and other experiments, one is forced to the conclusion that less than 20\% of the matter in the Universe is baryonic.

\subsection{Primordial Black Holes}

By the late 1990s, it had become clear that baryonic dark matter does not constitute a large fraction of the Universe's dark matter. Although these results seem to imply that the dark matter must consist of one or more new particle species, there remains a caveat to this conclusion: the dark matter might instead consist of black holes that formed before the epoch of Big Bang nucleosynthesis and with masses below the sensitivity range of microlensing surveys.

The possibility that black holes may have formed in the early Universe was discussed by Barnard Carr and Stephen Hawking as early as 1974~\cite{1974MNRAS.168..399C}. Such primordial black holes exhibit a characteristic mass that is on the order of the mass contained within the horizon at the time of formation, $M_{\rm horizon} \sim 10^{15} \, {\rm kg} \, \times \, (10^7 \, {\rm GeV}/T)^2$, allowing for a very large range of possible masses. A lower limit on this mass range can be placed, however, from the lack of Hawking-radiated gamma-rays from a primordial black hole population~\cite{1976ApJ...206....1P,MacGibbon:1987my}. Combining gamma-ray constraints~\cite{1998PhRvD..58j7502Y,1999PhRvD..59f3004K} with the null results of microlensing surveys yields an acceptable mass range of $10^{14} \, {\rm kg}$ to $10^{23} \, {\rm kg}$ for dark matter in the form of primordial black holes.

A major factor that has tempered the enthusiasm for primordial black hole dark matter pertains to the number of such objects that are expected to have formed in the early Universe. If one assumes an approximately scale-invariant spectrum of density fluctuations (normalized to that observed at large scales), the predicted formation rate is cosmologically negligible. To generate a relevant abundance of such black holes, one must postulate a large degree of non-gaussianity or other such features in the primordial power spectrum.

\section{Modified Gravity}
\label{modgrav}

In February of 1982, Mordehai Milgrom submitted a trio papers to the Astrophysical Journal~\cite{Milgrom:1983ca,Milgrom:1983pn,Milgrom:1983zz}. These papers, which Milgrom developed at the Weizmann Institute in Israel and while on sabbatical at Princeton's Institute of Advanced Study, provided the foundation for what would become the leading alternative to dark matter. This proposal, known as Modified Newtonian Dynamics, or MOND, was a seemingly simple one, but with extremely far reaching consequences. At the heart of MOND is the recognition that if instead of obeying Newton's second law, $F=ma$,  the force due to gravity scaled as $F = m a^2/a_0$ in the limit of very low accelerations ($a \ll a_0 \sim 1.2 \times 10^{-10}$ m/s$^2$), then it would be possible to account for the observed motions of stars and gas within galaxies without postulating the presence of any dark or otherwise hidden matter. In Milgrom's proposal, there was no dark matter. Instead, what astronomers had discovered was evidence of a new framework for gravity and dynamics, beyond that described by Newtonian physics or even by general relativity.

\subsection{Toward a Realistic Theory of MOND} 

Milgrom's initial proposal was not intended to represent a realistic theory, but rather was presented as the approximate weak-field limit of some unknown, but more complete framework. In its original form, it was not even clear whether MOND was merely a modification of the behavior of gravity, or was instead a more general correction to Newton's second law, applicable to all forces. Within the context of either interpretation, however, it has proven challenging to embed MOND-like behavior within a realistic theoretical framework. First of all, in its original formalism, MOND does not conserve momentum, angular momentum, or energy. Furthermore, Milgrom did not initially propose any means by which MOND could be embedded within a theory consistent with general relativity. Before MOND could be considered a viable alternative to dark matter, a more realistic version of this theory would have to be developed. And while significant progress has been made toward this goal over the past three decades, this progress has often been accomplished at great expense in terms of economy and simplicity. 

A first step in this direction was made in 1984 through the collaboration of Milgrom with Jacob Bekenstein, and their proposal of the AQUAdratic Lagrangian theory (AQUAL)~\cite{Bekenstein:1984tv}. In AQUAL, Bekenstein and Milgrom began with a modification of the Lagrangian of Newtonian gravity, rather than with a modification of Newton's second law. As a result, this approach automatically preserves the conservation of momentum, angular momentum, and energy, and respects the weak equivalence principle. And while the predictions of AQUAL are identical to those of MOND only in special and highly symmetric cases, the differences between the predictions of these two theories are typically modest (at the $\sim$$10\%$ level)~\cite{Milgrom:1986ib}.

Despite its advantages over the original version of MOND, AQUAL was still a modification of Newtonian dynamics, and is not compatible with the general theory of relativity. In order for any variation of MOND to be taken seriously, it would need to be able to account for the many varieties of relativistic phenomena that have been observed, including those of gravitational lensing and cosmological expansion. The first attempts to embed MOND into a relativistic framework involved theories with more than one metric. In relativistic AQUAL (RAQUAL)~\cite{Bekenstein:1984tv}, for example, the dynamics of matter and radiation are dictated by a metric that is distinct from the standard spacetime metric that applies to the gravitational field. The difference between these two metrics is the result of the presence of an additional scalar field which, along with matter, contributes to the gravitational potential. In this respect, RAQUAL shares some of the features of much earlier scalar-tensor theories of gravity~\cite{jordan,bransdicke}. To avoid causal problems resulting from the superluminal motion of the scalar field, however, RAQUAL had to be further modified~\cite{Bekenstein:1988zy}, and these changes were to the detriment of the theory's consistency with precision solar system tests. Even more problematic was the fact that these early attempts at a relativistic theory of MOND failed to adequately describe the phenomena of gravitational lensing. 
 
Although the deflection of light is predicted to occur in RAQUAL and other relativistic formulations of MOND, the magnitude of such lensing is generally expected to be proportional to the amount of (baryonic) mass that is present in the deflecting system.  In contrast, the degree of lensing that is observed around galaxy clusters is much larger than can be accounted for by the mass of the baryons alone. In this respect, RAQUAL cannot address the dark matter problem on cluster scales. 

Although other efforts to resolve this issue were attempted~\cite{Bekenstein:1993fs,Sanders:1996wk}, it was not until 2004 that Bekenstein proposed the first realistic solution to the problem of gravitational lensing in relativistic theories of MOND~\cite{Bekenstein:2004ne}. Since its proposal, Bekenstein's TeVeS theory -- short for Tensor-Vector-Scalar gravity -- has become the leading theory of MOND, and has received a great deal of attention. Beyond those of general relativity, TeVeS contains two additional fields, three free parameters, and one free function. On the one hand, this freedom makes TeVeS somewhat limited in its predictive power. On the other, it provides TeVeS with enough flexibility to potentially be consistent with gravitational lensing observations and other cosmological considerations, such as those pertaining to structure formation and the cosmic microwave background.

\subsection{Observational Successes and Failures}

Early in its history, it was appreciated that MOND was capable of explaining the observed dynamics of many spiral and elliptical galaxies. MOND also, however, made predictions for the behavior of low surface brightness galaxies, whose dynamics had not yet been well measured. The fact that such systems were later found to be compatible with MOND~\cite{Casertano:1991sh,McGaugh:1998tq} served to bolster interest in the theory. Today, MOND appears to be compatible with the observed rotations curves of hundreds of spiral galaxies~\cite{Begeman:1991iy,Sanders:1996ua,Milgrom:2004ba,Milgrom:2006xn}. 

%Considering that MOND -- in the weak-field limit -- requires only one free parameter, $a_0$, this is seen by many proponents as an impressive accomplishment. 

In addition to galactic rotation curves, MOND also provides an explanation for the empirical Tully-Fisher formula~\cite{Tully:1977fu}, which relates the intrinsic luminosities and rotational velocities of spiral galaxies, $L \propto V_{\rm rot}^\alpha$, where $\alpha\approx4$. If one assumes a common mass-to-light ratio for all galaxies, MOND predicts precisely this relationship, with a value of $\alpha=4$, which is consistent with observations~\cite{McGaugh:2000sr}.

%.At the time of MOND's original proposal, the value of this exponent had not yet been definitely measured The fact that observations have since converged to a value consistent with the prediction of MOND~\cite{McGaugh:2000sr,McGaugh:2005qe} is another success of the theory. Whether the observed Tully-Fisher relationship should be expected within the context of $\Lambda$CDM cosmology is less clear, as it depends on the complex interplay between dark matter and baryonic processes.  

On the scale of galaxy clusters, MOND has not been nearly as successful. While MOND does reduce the need for additional mass in clusters, significant quantities of dark matter are still required. If the three known species of neutrinos were as heavy as $m_{\nu}\sim$1-2 eV (near the upper limits from beta decay experiments), it has been suggested that they might be able to account for this discrepancy, essentially acting as dark matter in clusters~\cite{clusterneutrinos,Angus:2006ev,Sanders:2007zn}. Massive neutrinos can also help to reduce to some degree the discrepancy between measurements of the cosmic microwave background and the predictions of TeVeS~\cite{Skordis:2005xk}.

%%%%%

In recent years, the debate over MOND has been focused on the use of gravitational lensing to measure the mass profiles of galaxy clusters. The idea that lensing could be used to determine the mass of a galaxy or a galaxy cluster was first proposed by Fritz Zwicky in his famous paper of 1937 (see Chapter~\ref{zwickysmith}). It was more than 40 years later that the first gravitational lens was observed~\cite{1979Natur.279..381W} -- two mirror images of a quasar -- and another decade after that before the first observations were made of lensing by a galaxy cluster~\cite{1986BAAS...18R1014L,1987A&A...172L..14S}. Today, gravitational lensing is frequently used to study the properties of clusters (see e.g. Refs.~\cite{2013SSRv..177...75H, 2010pdmo.book...56M} for recent reviews). 

\begin{figure}[t]
\centering
\includegraphics[width=0.9\textwidth]{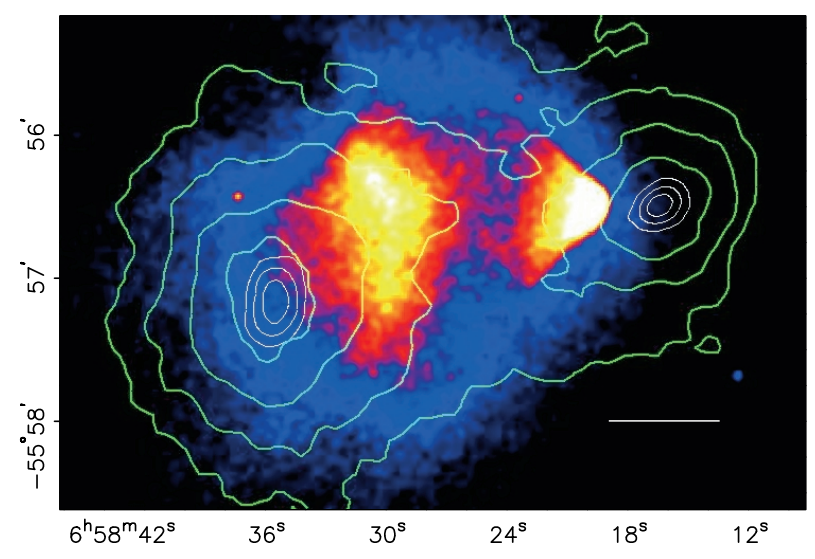}
\caption{The bullet cluster. The colored map represents the X-ray image of this system of merging clusters, obtained in a 500 second exposure with Chandra. The white bar is shown for scale, and represents a distance of 200 kpc at the location of the cluster. The green contours denote the reconstructed lensing signal, proportional to the projected mass in the system. From Ref.~\cite{2006ApJ...648L.109C}.\label{fig:bullet}}
%\end{center}
\end{figure}

In 2006, a group of astronomers including Douglas Clowe transformed the debate between dark matter and MOND with the publication of an article entitled, ``A direct empirical proof of the existence of dark matter''. In this paper, the authors described the observations of a pair of merging clusters collectively known as the ``bullet cluster'' (and also known as 1E0657-558)~\cite{2006ApJ...648L.109C}. As a result of the clusters' recent collision, the distribution of stars and galaxies is spatially separated from the hot X-ray emitting gas (which constitutes the majority of the baryonic mass in this system). A comparison of the weak lensing and X-ray maps of the bullet cluster clearly reveals that the mass in this system does not trace the distribution of baryons (see Fig.~\ref{fig:bullet}).  Another source of gravitational potential, such as that provided by dark matter, must instead dominate the mass of this system. 
 
Following these observations of the bullet cluster (and of other similar systems), many researchers expected that this would effectively bring the MOND hypothesis to an end. In the years since, however, anything but has taken place. Since the introduction of TeVeS, MOND has continued to attract a great deal of attention, despite its failure to address the dynamics of galaxy clusters, and in particular the bullet cluster. In addition to massive neutrinos, some authors have considered the possibility that TeVeS's vector field might source the gravitational potential of the bullet cluster, itself acting much like dark matter on cluster scales. Similarly, the failure of TeVeS to predict the observed ratio of the second and third peaks of the cosmic microwave background's angular power spectrum might be plausibly averted if some of TeVeS's additional degrees-of-freedom behaved much like cold dark matter during the early history of the Universe. And although this possibility goes somewhat against the original spirit of MOND, it is hard to rule out at this time. Taken together, the bullet cluster and other increasingly precise cosmological measurements have been difficult to reconcile with all proposed versions of MOND, and it remains unclear whether TeVeS, in some form, might be compatible with these observations~\cite{Skordis:2005xk,Dodelson:2006zt,Skordis:2008pq}. For reviews, See Refs.~\cite{Skordis:2009bf,Famaey:2011kh}.

\section{Piecing the puzzle}
\label{chap:cosmology}

\subsection {Discrepancies At All Scales}

When Fritz Zwicky proposed in 1933 that dark matter might be responsible for the high velocity dispersion of galaxies in the Coma Cluster (see Chapter~\ref{zwickysmith}), he was familiar with the concept of dark matter, and with earlier attempts to dynamically measure the density of dark matter in the Galaxy. Over the decades that followed, however, the presence of dark matter in clusters and in galaxies were discussed largely independently of each other. It wasn't until the 1960s that mass discrepancies on multiple scales began again to be considered within a common context.

%Arrigo Finzi was perhaps the first, in 1963, to attempt a coherent discussion of the large velocity dispersions in clusters of galaxies, the surprising rotation curve of M31, and the mass determinations of the Milky Way \cite{1963MNRAS.127...21F}. 

%identifying discrepancies in astrophysical observations on a wide range of scales, or perhaps just because of that. 

In his pioneering paper of 1963, Arrigo Finzi cited Zwicky's 1933 work on galaxy clusters, the 1957 observation of M31's rotation curve from van de Hulst et al., as well as more recent determinations of the mass of the Milky Way, and argued in favor of a common interpretation for these phenomena~\cite{1963MNRAS.127...21F}.  He then went on to consider various possible forms of what we would today call ``baryonic'' dark matter, ruling them out one-by-one. He even went as far as to suggest that these phenomena might be explained by modifying Newton's gravitational force law, so that it scaled as $r^{-3/2}$ at large distances. 

Despite the highly original and prescient nature of Finzi's work, it was largely ignored by the scientific community~\cite{2010dmp..book.....S}, attracting only 50 citations over the past 50 years. Although it is impossible to unambiguously identify the precise reasons for this, the very bold nature of Finzi's conclusions may have been difficult for many of his colleagues to accept, or even seriously consider. In any case, this work had little impact, and it would be another decade before other scientists began to pursue similar lines of inquiry. 

%this line of inquiry would not be revisited by others for another ten years.  

As mentioned in Chapter~\ref{Chap:rotcurves}, two independent groups published groundbreaking papers in 1974, each presenting a strong case for the existence of large amounts of mass in the outer parts of galaxies. The first of these papers, by the Estonian astronomers Jaan Einasto, Ants Kaasik and Enn Saar,  was submitted on April 10 and was entitled ``Dynamic evidence on massive coronas of galaxies"~\cite{1974Natur.250..309E}. These authors began with a discussion of galactic rotation curves, citing the work of Roberts that would be eventually published in Ref.~\cite{1975gaun.book..309R}, and presented an analysis of rotation curve data that included estimates for the contributions from stars for five galaxies of different mass. They argued that the discrepancy between the total mass and the stellar mass implied the existence of a ``corona'', consisting of a ``previously unrecognised, massive population''. They then used 105 pairs of galaxies to estimate the total mass and dimensions of their galactic coronas, concluding that the total mass of galaxies exceeded that in stars by an order of magnitude. Finally, the authors argued that these new mass estimates could also explain the mass discrepancy that had been observed in clusters. Similar arguments had also been sketched earlier by the Einasto, including at the 1972 IAU meeting in Athens~\cite{1972TarOT..40....3E}.

On May 28 -- about six weeks after Einasto et al. -- Jerry Ostriker, Jim Peebles and Amos Yahil submitted a paper of similar content and scope, entitled ``The size and mass of galaxies, and the mass of the universe"~\cite{1974ApJ...193L...1O}. This paper did not present any new observations, but instead compiled existing estimates for the masses of (mostly giant spiral) galaxies. They begin with galactic rotation curves, citing the papers of Roberts and Rots~\cite{1973A&A....26..483R} and Rogstad and Shostak~\cite{1972ApJ...176..315R} as evidence for their flatness in the outer parts of galaxies. The authors then went on to build a case for the existence of large amounts of dark matter in the outer parts of galaxies, based on mass estimates from galaxy pairs, the dynamics of dwarf galaxies, and the so-called timing argument for the Local Group. And although the observations presented in this paper were not new, and were subject to large uncertainties, the authors appear to have been confident in their conclusions, stating that the trend of increasing mass with increasing radius is ``almost certainly real'', and arguing that this trend was in line with the ``virial discrepancy'' that had been observed in clusters and groups of galaxies~\cite{1970ApJ...162..411R,1971ApJ...170..199F,1974ApJ...188..451R}. The first sentences of this paper's body summarizes well the sentiment of the authors:
\begin{quote} 
{\it ``There are reasons, increasing in number and quality, to believe that the masses of ordinary galaxies may have been underestimated by a factor of 10 or more. Since the mean density of the Universe is computed by multiplying the observed number density of galaxies by the typical mass per galaxy, the mean density of the Universe would have been underestimated by the same factor."}
\end{quote}

In 1979, Sandra Faber and John Gallagher published an influential review, ``Masses and mass-to-light ratios of galaxies"~\cite{1979ARA&A..17..135F}, which played an important role in crystallizing the opinion among cosmologists and astronomers that dark matter was indeed abundant in the Universe. Interestingly, they chose not to use the terms ``corona'' or ``halo'', as suggested by the two above mentioned papers, but instead adopted the phrase ``massive envelope'' to describe the distribution of dark matter in astrophysical systems\footnote{In the discussions that took place as part of our research for this historical review, we encountered a considerable range of opinions regarding the relative importance of galactic rotation curves in establishing the existence of dark matter. This supports a picture in which different groups of scientists found quite different lines of evidence to be compelling during this period of time. Despite these disagreements regarding the strengths and weaknesses of the various observations and arguments, a consensus nonetheless began to emerge in favor of dark matter's existence.}.

%Although there was little information at that time pertaining to the spatial distribution of dark matter, important clues about the shape and density profile were about to arise as the result of a new field of research: cosmological computer simulations. 

\subsection{Cosmology}

As astronomers continued to gather information on the masses of galaxies, and on other observables of cosmological relevance, cosmologists began to increasingly reflect upon the implications of those findings for the formation of structure and the evolution of the Universe. In 1974, the same year as the two key papers described above~\cite{1974Natur.250..309E,1974ApJ...193L...1O}, Richard Gott, James Gunn, David Schramm and Beatrice Tinsley published a paper that provides us with an illuminating snapshot of the status of cosmology at that time~\cite{1974ApJ...194..543G}. The conclusions of this paper, entitled ``An unbound universe", appear within the original abstract:
\begin{quote}
{\it ``A variety of arguments strongly suggest that the density of the universe is no more than a tenth of the value required for closure. Loopholes in this reasoning may exist, but if so, they are primordial and invisible, or perhaps just black."}
\end{quote}
In this paper, the authors argued that the body of astronomical data indicated that there was simply not enough matter in the Universe -- even accounting for the large mass-to-light ratios observed among galaxies -- to equal or exceed the critical density of the Universe. Among other caveats to this conclusion, they considered possible contributions from low-mass neutrinos, as had been suggested by Cowsik and McClelland, but ultimately ruled out this possibility as well. 

In the early 1980s, the introduction of the theory of inflation profoundly changed the thinking of the cosmological community, and allowed one for the first time to make specific predictions for the total cosmological density and for the spectrum of density perturbations~\cite{1981PhRvD..23..347G,1982PhLB..108..389L,1982PhLB..117..175S,1982PhLB..115..295H,1982PhLB..115..295H,1982PhRvL..49.1110G,1983PhRvD..28..679B}. This began a decade long struggle to reconcile models of structure formation with what had by then become the ``theoretical imperative'' of a flat Universe~\cite{firstcdm}. This struggle was exacerbated by estimates of the cosmological matter density arising from galaxy clusters which pointed toward a total abundance of matter -- including dark matter, by then accepted by most cosmologists -- that was clearly insufficient to close the Universe~\cite{1993Natur.366..429W}. The resolution to this problem had to await the discovery of the accelerating expansion rate of the Universe, and the contribution to the total energy density arising from a cosmological constant or ``dark energy''.

Meanwhile, Jim Peebles had pointed out that the absence of fluctuations in the cosmic microwave background at a level of $\sim$$10^{-4}$ was incompatible with a Universe that was composed of only baryonic matter, and argued that this problem would be relieved if the Universe was instead dominated by massive, weakly interacting particles, whose density fluctuations could begin to grow prior to decoupling~\cite{1982ApJ...263L...1P} (see also, Ref.~\cite{1981SvA....25...14C}). This and other papers that will be discussed in the next section received enormous attention from the scientific community, and rapidly led to the establishment of cold dark matter as the leading paradigm to describe the structure and evolution of the Universe at all scales. 

\subsection{Numerical Simulations}
\label{sec:simulations}

Much of our current understanding of the structure and evolution of dark matter halos in the Universe is based on the results of computer simulations. Such explorations have a longer history than one might expect. Working in the 1940s, the ingenious Swedish scientist Erik Holmberg\index{Holmberg, Erik|(} exploited the fact that light follows the same inverse square law as the gravitational force, and performed the first simulation of the interaction between two galaxies on an analog computer that consisted of 74 light-bulbs, photo-cells and galvanometers. He then calculated the amount of light received by each cell, and manually moved the light bulbs in the direction that received the most light. 

Holmberg published his paper in November of 1941, shortly before the United States entered World War II. In the following years, the work of many research institutes ground to a halt, but science meanwhile continued to make progress, thanks in large part to the enormous resources made available to military research, especially at the Los Alamos National Laboratory, at which computers and advanced numerical techniques were developed within the context of the Manhattan Project. The first application of such computers to gravitational systems was arguably performed by John Pasta  and Stanislaw Ulam in 1953. Their numerical experiments were performed on the Los Alamos computer, which by then had been applied to a variety of other problems, including early attempts to decode DNA sequences and the first chess-playing program. A number of other pioneering studies exploring the evolution of a system of gravitationally interacting massive particles appeared throughout the 1960s and 1970s, gradually increasing the number of simulated particles from $N \sim 100$ in the early works to $N \sim 1000$~\cite{1960ZA.....50..184V,1963ZA.....57...47V,1963MNRAS.126..223A,1970AJ.....75...13P,1976MNRAS.177..717W}. 

%Twenty years later it was at another american agency, NASA, that the most powerful computers were running, thanks to the extraordinary space program that culminated with the landing of the {\it Apollo 11} mission on the Moon. Two young astrophysicists, Alar and Juri| Toomre, had access in the early 1970s to one of only two NASA's IBM 360-95 computers, complete with high resolution graphics workstations and auxiliary graphics rendering machines --- computing facilities far in advance of any other astrophysics laboratory.  They set up a series of simulations of galaxy grazings and collisions with a simple code that described the galaxies as two massive points surrounded by a disk of test particles. The outcome of this analysis was a very influential paper, published in 1972, that contained a detailed discussion of the role of collisions in the formation of galaxies. 

By the early 1970s, it had become possible to numerically simulate the dynamics of galaxies. Simulations carried out by Richard Miller, Kevin Prendergast and William Quirk~\cite{1970ApJ...161..903M} as well as by Frank Hohl~\cite{1971ApJ...168..343H} each found rotationally suppored galaxies consisting of a stellar disk to be unstable, in contradition with observations. Instead of reaching an equilibrium configuration, such systems were found to change rapidly, forming bars and evolving toward a more elliptical and pressure supported configuration. The solution to this problem was proposed in 1973 by Jerry Ostriker and Jim Peebles, who recognized that a rotationally supported stellar disk could be stable if embedded within a massive spherical halo~\cite{1973ApJ...186..467O}.

The first attempt to numerically solve the formation and evolution of cosmological structures in an expanding universe was presented in a famous paper published in 1974 by William Press and Paul Schechter~\cite{1974ApJ...187..425P}. This was followed by a number of developments in the late 1970s and early 1980s that significantly advanced the power of such endeavors (see, for example, Refs.~\cite{1979ApJ...228..664A,1979ApJ...228..684T,1979ApJ...234...13G,1983MNRAS.204..891K,1983Natur.305..196C,1981MNRAS.194..503E,1983million}). First, 
%
%whereas the simulations carried out in the 1970s could follow the trajectories of up to $\sim$$10^3$ particles, 
%
a combination of improvements in processor speed and in numerical techniques made it possible for the first time to simultaneously simulate millions of particles. Second, the newly proposed theory of inflation~\cite{Guth:1980zm,Linde:1981mu} offered a physical means by which initial density perturbations could be generated, providing the initial conditions for cosmological simulations. And third, the results of the first large 3D survey of galaxies (the CfA redshift survey) were published in 1982, providing a distribution that could be directly compared with the output of simulations. 

In some ways, the results of cosmological simulations do not depend much on what the dark matter consists of. In particular, they are largely insensitive to the electroweak or other non-gravitational interactions that may (or may not) be experienced by dark matter particles -- for the purposes of structure formation, such particles are effectively ``collisionless''. What does impact the results of such simulations, however, is the initial velocity distribution of the dark matter particles~\cite{Doroshkevich:1980zs,1980PhRvL..45.1980B,1983ApJ...274..443B}. Importantly, this provides cosmologists with a way to discriminate between different classes of dark matter candidates. Standard model neutrinos, for example, decoupled from thermal equilibrium in the early Universe at a temperature that is much greater than their mass, and thus remained highly relativistic throughout the epoch of structure formation. In contrast, supersymmetric neutralinos are predicted to freeze-out of thermal equilibrium at a temperature below their mass, and are thus non-relativistic throughout cosmic history. Axions generated through misalignment production are also predicted to be produced with non-relativistic velocities.

At the largest scales -- those associated with galaxy clusters and superclusters -- cosmological simulations predict a pattern of structure that is largely insensitive to the initial velocities of the dark matter. At smaller scales, however, density fluctuations can be washed out by the random thermal motion of individual dark matter particles. As a result, the growth of small scale structure is predicted to be suppressed if the dark matter is relativistic, or ``hot''~\cite{1981ApJ...243....1S,1982ApJ...258..415P}.
%
%In simulations of such relativistic, or `hot' dark matter, the growth of all structures smaller than the largest galaxy clusters is suppressed, leading superclusters (with masses of $\sim10^{15}\, M_{\odot}$) to form first. In a Universe dominated by hot dark matter, the formation of galaxies was delayed until much later in cosmic history, and then only as a result of the fragmentation and breaking up of larger halos.
%
Non-relativistic, or ``cold'' dark matter particles undergo a very different sequence of structure formation. The much shorter free-streaming length of such particles allow them to form very low mass halos; roughly in the range of $\sim$$10^{-3}\,$ to $\sim$$10^{-9} M_{\odot}$ for a typical neutralino, for example. These very small halos form very early in the Universe's history, and then go on to merge with one another, gradually building up larger and larger dark matter structures. This bottom-up, or hierarchical, process of structure formation is in stark contrast to the top-down sequence predicted for hot dark matter.

Simulations of large scale structure are, of course, only useful if their results can be compared to the actual patterns of structure found in the Universe. This was made possible with the CfA survey, which was the first extensive 3D survey of galaxies in the local Universe~\cite{cfa}. Among other features, CfA revealed the first indications of the ``cosmic web'', which described the distribution of matter on the largest scales. This survey also identified the presence of significant structure on sub-cluster scales, in conflict with the predictions of hot dark matter simulations~\cite{hdmfails}. 

In the wake of the failures of hot dark matter, it was quickly becoming appreciated that cold dark matter could do a much better job of accounting for the observed patterns of large scale structure. To quote the 1984 paper by George Blumenthal, Sandra Faber, Joel Primack, and Martin Rees~\cite{blumenthal}:
\begin{quote}
{\it ``We have shown that a universe with $\sim$10 times as much cold dark matter as baryonic matter provides a remarkably good fit to the observed universe. This model predicts roughly the observed mass range of galaxies, the dissipational nature of galaxy collapse, and the observed Faber-Jackson and Tully-Fisher relations. It also gives dissipationless galactic halos and clusters. In addition, it may also provide natural explanations for galaxy-environment correlations and for the differences in angular momenta between ellipticals and spiral galaxies."}
\end{quote}

The first simulations of cold dark matter were carried out by Marc Davis, George Efstathiou, Carlos Frenk, and Simon White, who published their results in 1985~\cite{firstcdm}. The resemblance of their simulated distribution of dark matter halos to that of the galaxies in the CfA survey was clear, serving to further elevate the status of cold dark matter within the cosmological community.

%The successes of cold dark matter became only more clear in light of the first cold dark matter simulations, which were published in 1985~\cite{firstcdm}. The resemblance to the distribution of galaxies in the CfA survey was clear.

By middle of the 1980s, the paradigm of cold dark matter was well on its way to becoming firmly established. And although scenarios involving mixed dark matter (containing significant quantities of both cold and hot dark matter) and warm dark matter (suppressing structure only on the scale of dwarf galaxies and below) would each continue to be discussed in the literature, the possibility that the dark matter was dominated by neutrinos or other relativistic particles was quickly abandoned.

A decade later, the predictions of cosmological simulations had shifted in focus from the distribution of cold dark matter halos to the shapes of those halos. In 1996, Julio Navarro, Carlos Frenk and Simon White published a remarkable result, based on an analysis of the halos generated in their high-resolution cold dark matter simulations~\cite{1996ApJ...462..563N}:
\begin{quote}
{\it The spherically averaged density profiles of all our halos can be fit over two decades in radius by scaling a simple universal profile. The characteristic overdensity of a halo, or
equivalently its concentration, correlates strongly with halo mass in a way which reflects the mass dependence of the epoch of halo formation.}
\end{quote}

The simple fitting formula derived by the authors became known as the Navarro-Frenk-White profile. This parametrization is still widely used today, and represents the primary benchmark for most dark matter detection studies, despite the fact that it is expected to be inaccurate in the innermost regions of galaxies, where baryons dominate the gravitational potential.

In more recent years, the frontier for cosmological simulations has focused on the implementation of baryonic physics, including the hydroynamical evolution of gas in astrophysical structures, stellar formation, and feedback from supernova explosions and black holes.  Current simulations are not yet able to resolve all relevant scales -- which range between sub-parsec distances for stellar formation to Gpc scales for cosmological structures -- but implement baryonic physics through the introduction of suitable ``sub-grid'' parameters which attempt to encode the collective behaviour of large amounts of gas and stars. Such parameters are generally tuned to match observable quantities, such as the galaxy mass function and the galaxy-central black hole mass relation, as in e.g. the recent suite of Eagle simulations~\cite{2015MNRAS.446..521S}.
 
\section{The Hunt for Dark Matter Particles}
\label{sec:searches}

As particle physicists became increasingly interested in the problem of the missing matter of the Universe, some began to turn their attention toward ways that individual particles of dark matter might be detected, either directly or indirectly. Although many of the leading techniques were first conceived of in the 1980s, dark matter searches have continued with vigor ever since, occupying the attentions of generations of experimental particle-astrophysicists.

\subsection{Scattering with Nuclei}

In 1984, an article by Andrzej Drukier and Leo Stodolsky at the Max Planck Institute in Munich appeared in {\it Physical Review D}, discussing techniques that might be used to detect neutrinos scattering elastically off nuclei~\cite{Drukier:1983gj}. Among other possibilities, the article proposed the use of a superconducting colloid detector, consisting of micron-scale superconducting grains maintained at a temperature just below their superconducting transition. Even a very small quantity of energy deposited by the recoil of an incident neutrino could cause a superconducting grain to flip into the normal state, collapsing the magnetic field and producing a potentially measurable electromagnetic signal. In January 1985, Mark Goodman and Ed Witten submitted a paper to the same journal, arguing that this technology could also be used to detect some types of dark matter particles~\cite{Goodman:1984dc}\footnote{A similar paper by Ira Wasserman~\cite{Wasserman:1986hh} was submitted shortly after Goodman and Witten's.}. Although Drukier and Stodolsky's original detector concept was never employed at a scale sensitive to dark matter, the broader notion of experiments capable of detecting $\sim$1-100 keV nuclear recoils provided a path through which it appeared possible to test the WIMP hypothesis. 

In their original paper, Goodman and Witten considered three classes of dark matter candidates:~1) those that undergo coherent scattering with nuclei (also known as spin-independent scattering), 2) those that scatter with nuclei through spin-dependent couplings, and 3) those with strong interactions.  The first two of these three categories provide the basis for how most direct dark matter detection results have since been presented. If mediated by unsuppressed couplings to the $Z$ boson (an important early benchmark), coherent scattering was predicted to lead to large scattering rates, typically hundreds or thousands of events per day per kilogram of target material. With such high rates, the prospects for detecting dark matter in the form of a heavy neutrino or sneutrino appeared very encouraging. Dark matter candidates that scatter with nuclei only through spin-dependent couplings, in contrast, were generally predicted to yield significantly lower rates, and would require larger and more sensitive detectors to test. Even as early as in this first paper, Goodman and Witten pointed out that such experiments would have difficultly detecting dark matter particles lighter than $\sim$1-2 GeV, due to the modest quantity of momentum that would be transferred in the collisions.

The first experiment to place constraints on the scattering cross section of dark matter with nuclei was carried out in 1986 at the Homestake Mine in South Dakota by a collaboration of scientists at the Pacific Northwest National Laboratory, the University of South Carolina, Boston University, and Harvard~\cite{Ahlen:1987mn}. Using a low-background germanium ionization detector (originally designed to search for neutrinoless double beta decay), they accumulated an exposure of 33 kg-days, yielding a limit that significantly constrained dark matter candidates with unsuppressed spin-independent scattering cross sections with nuclei (such as heavy neutrinos or sneutrinos)~\cite{Ahlen:1987mn}. Shortly thereafter, similar results were obtained by an independent collaboration of scientists from the Universities of California at Santa Barbara and Berkeley~\cite{Caldwell:1988su}. 

Despite the importance of these first dark matter scattering limits, the reach of such detectors quickly became limited by their backgrounds, making it difficult to achieve significant improvements in sensitivity. One possible solution to this problem, first suggested by Andrzej Drukier, Katherine Freese, and David Spergel~\cite{Drukier:1986tm}, was to search for an annual variation in the rate of dark matter induced events in such an experiment, as was predicted to result from the combination of the Earth's motion around the Sun and the Sun's motion through the dark matter halo. Such a technique could, in principle, be used to identify a signal of dark matter scattering over a large rate of otherwise indistinguishable background events. The most well known group to employ this technique was the DAMA/NaI Collaboration (and later DAMA/LIBRA). The original DAMA/NaI experiment consisted of nine 9.70 kg scintillating thallium-doped sodium iodide crystals, located in Italy's deep underground Gran Sasso Laboratory. In 1998, they published their first results, reporting the observation of an annually modulating rate consistent with dark matter scattering~\cite{Bernabei:1998fta}. Over the past nearly two decades, DAMA's signal has persisted and become increasingly statistically significant as more data was collected~\cite{Bernabei:2003za}, including with the more recent DAMA/LIBRA detector~\cite{Bernabei:2008yi,Bernabei:2010mq}. At this point in time, it seems hard to reconcile dark matter interpretations of the DAMA/LIBRA signal with the null results of other direct detection experiments. On the other hand, no convincing alternative explanation for this signal has been so far identified.
% is unclear how to interpret these results. On the one hand, other experiments have been unable to detect a signal compatible with a dark matter interpretation of DAMA's modulation.  It is fair to say, however, that the scientific community is currently very skeptical of dark matter interpretations of the DAMA result.

\begin{figure}
    \begin{center}
        \includegraphics[width=0.85\textwidth]{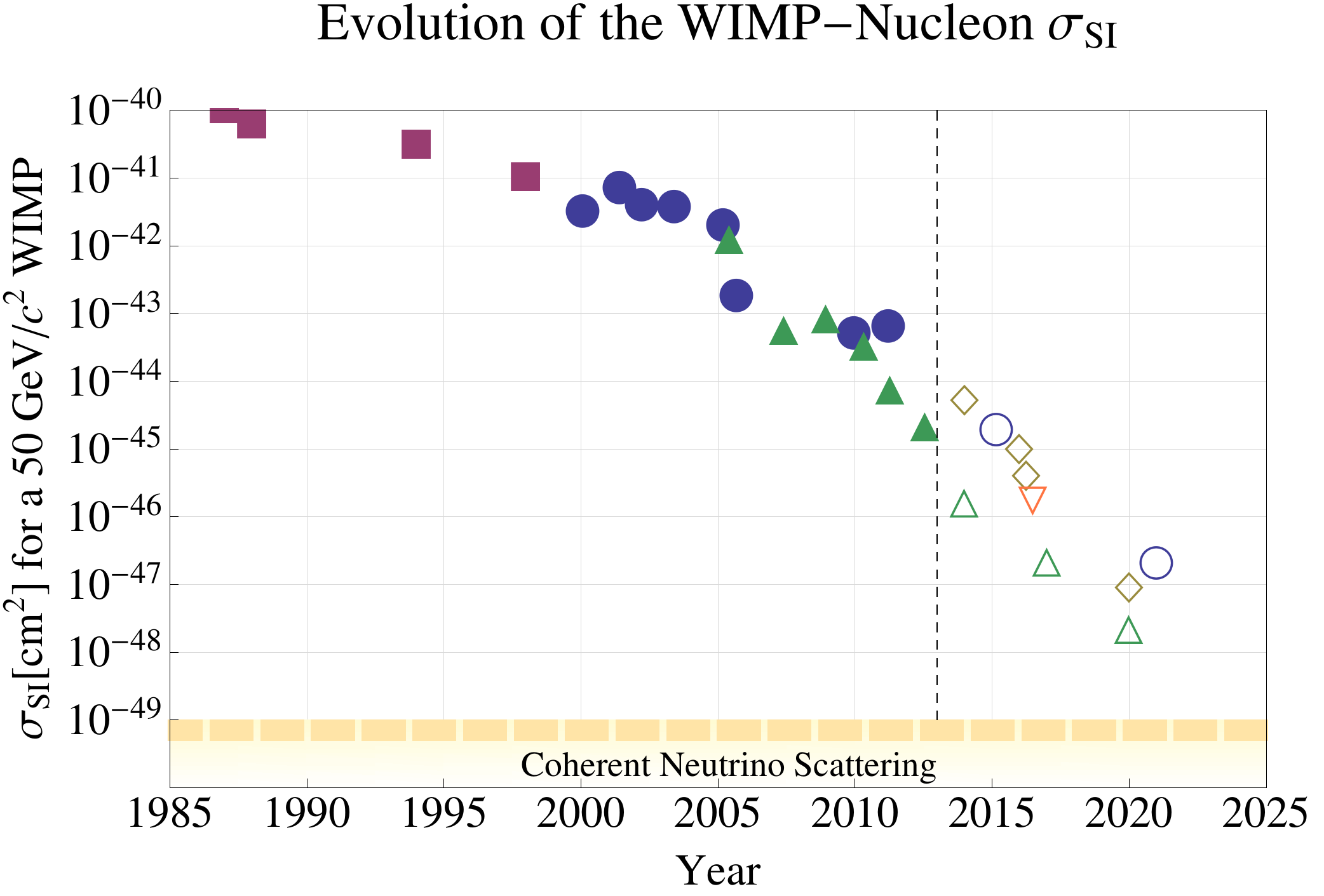}
    \caption{The past and projected evolution of the spin-independent WIMP-nucleon cross section limits for a 50 GeV dark matter particle. The shapes correspond to limits obtained using different detectors technologies: cryogenic solid state detectors (blue circles), crystal detectors (purple squares), liquid argon detectors (brown diamonds), liquid xenon detectors (green triangles), and threshold detectors (orange inverted triangle). Taken from Ref.~\cite{Cushman:2013zza}.}
    \label{fig:direct}
    \end{center}
\end{figure}

During the period of time that DAMA/NaI was being developed and collecting its first data, experimental techniques were being pursued that could discriminate dark matter-like nuclear recoil events from various backgrounds. These efforts ultimately lead to the technologies employed by the CDMS (Cryogenic Dark Matter Search), EDELWEISS (Experience pour DEtecter Les Wimps En Site Souterrain), and CRESST (Cryogenic Rare Event Search with Superconducting Thermometers) Collaborations. These experiments each made use of two-channel detectors, capable of measuring both ionization and heat (CDMS, EDELWEISS) or scintillation and heat (CRESST), the ratio of which could be used to discriminate nuclear recoil events from electron recoils generated by gamma and beta backgrounds. All three of these experiments employed crystalline target materials, maintained at cryogenic temperatures, consisting of germanium and silicon, germanium, and calcium tungstate, respectively. Throughout most the first decade of the 21st century, the CDMS and EDELWEISS experiments lead the field of direct detection, providing the most stringent constraints and improving in sensitivity by more than two orders of magnitude over that period of time (see Fig.~\ref{fig:direct}).

In order to continue to increase the sensitivity of direct dark matter experiments, it was necessary for experiments to employ ever larger targets, gradually transitioning from the kilograms of detector material used by EDELWEISS and CDMS (9.3 kg in the case of SuperCDMS) to the ton-scale and beyond. Cryogenic solid state detectors, however, have proven to be costly to scale up into ton-scale experiments. In the late 1990s, Pio Picchi, Hanguo Wang and David Cline pioneered an alternative technique that exploited liquid noble targets (most notably liquid xenon). 
%Since 2010 or so, liquid xenon detectors have yielded the most stringent limits on dark matter scattering with nuclei. 
Like solid state detectors, such experiments discriminate nuclear recoils from electron recoils by measuring two quantities of deposited energy; in this case scintillation and ionization. Between 2010 and 2015, the XENON100 and LUX experiments (each of which utilize a liquid xenon target) have improved upon the limits placed by CDMS by approximately two orders of magnitude. It is generally anticipated that future experiments employing liquid xenon targets (XENON1T, LZ, XENON-NT) will continue along this trajectory for years to come. 

As CDMS, EDELWEISS, XENON100, LUX and other direct detection experiments have increased in sensitivity over the past decades, they have tested and ruled out an impressive range of particle dark matter models. And although results from the CoGeNT~\cite{Aalseth:2010vx,Aalseth:2011wp}, CRESST~\cite{Angloher:2011uu}, and CDMS~\cite{Agnese:2013rvf} experiments were briefly interpreted as possible dark matter signals, they now appear to be the consequences of poorly understood backgrounds~\cite{Angloher:2014myn,cogentkelso} and/or statistical fluctuations. While many viable WIMP models remain beyond the current reach of this experimental program, a sizable fraction of the otherwise most attractive candidates have been excluded. Of particular note is the fact that these experiments now strongly constrain dark matter particles that scatter coherently with nuclei through Higgs exchange, representing an important theoretical benchmark.

\subsection{Annihilation and Decay}

In the 1978 Valentine's Day issue of {\it Physical Review Letters}, there appeared two articles that discussed -- for the first time -- the possibility that the annihilations of pairs of dark matter particles might produce an observable flux of gamma rays. And although each of these papers (by Jim Gunn, Ben Lee,\footnote{In regards to Ben Lee, who died in a traffic accident in 1977, this article was published posthumously.} Ian Lerche, David Schramm and Gary Steigman~\cite{Gunn:1978gr}, and by Floyd Stecker~\cite{Stecker:1978du}) focused on dark matter in the form of a heavy stable lepton (i.e.~a heavy neutrino), similar calculations would later be applied to a wide range of dark matter candidates. On that day, many hopeless romantics became destined to a lifetime of searching for signals of dark matter in the gamma-ray sky.

At the time, the most detailed measurement of the astrophysical gamma-ray background was that made using data from the Small Astronomy Satellite (SAS) 2~\cite{SAS2}. Although the intensity of 35-100 MeV gamma rays measured by this telescope ($\sim$$6 \times 10^{-5}$ cm$^{-2}$ s$^{-1}$ sr$^{-1}$) was several orders of magnitude higher than that predicted from annihilating dark matter particles smoothly distributed throughout the Universe, it was recognized that inhomogeneities in the dark matter distribution could increase this prediction considerably. In particular, annihilations taking place within high-density dark matter halos, such as that of the Milky Way, could plausibly produce a flux of gamma rays that was not much fainter than that observed at high galactic latitudes, and with a distinctive gradient on the sky~\cite{Gunn:1978gr,Stecker:1978du}. Focusing on GeV-scale dark matter particles, Gunn et al.~went as far as to state that such a signal ``may be discoverable in future $\gamma$-ray observations''. 

Several years later, in 1984, Joe Silk and Mark Srednicki built upon this strategy, considering not only gamma rays as signals of annihilating dark matter particles, but also cosmic-ray antiprotons and positrons~\cite{Silk:1984zy} (see also, Refs.~\cite{1985PhRvL..55.2622S,1988PhLB..214..403E,Kamionkowski:1990ty}). They argued that the observed flux of $\sim$0.6-1.2 GeV antiprotons~\cite{Buffington:1981zz} provided the greatest sensitivity to annihilating dark matter, and noted that $\sim$10 GeV WIMPs would be predicted to produce a quantity of cosmic-ray antiprotons that was comparable to the observed flux.

In 1985, Lawrence Krauss, Katherine Freese, David Spergel and William Press published a paper suggesting that neutrinos might be detected from dark matter annihilating in the core of the Sun~\cite{Krauss:1985ks} (see also, Ref.~\cite{Press:1985ug}). Shortly thereafter, Silk, Olive, and Srednicki pointed out that not only could elastic scattering cause dark matter particles to become gravitationally bound to and captured within the Sun, but that the number of WIMPs captured over the age of the Solar System could be sufficiently high to attain equilibrium between the processes of capture and annihilation~\cite{Silk:1985ax}. Observations over the subsequent few years by the proton decay experiments IMB, FREJUS, and Kamioka capitalized on this strategy, strongly constraining some classes of dark matter candidates, most notably including light electron or muon sneutrinos. Similar approaches using dark matter capture by the Earth were also proposed around the same time~\cite{Krauss:1985aaa,Freese:1985qw}. 

In the decades that followed, measurements of astrophysical gamma ray, antimatter, and neutrino fluxes improved dramatically. In parallel, the scientific community's understanding of the astrophysical sources and propagation of such particles also matured considerably. Information from successive gamma-ray satellite missions, including COS-B~\cite{Hermsen:1986bc}, EGRET (Energetic Gamma Ray Experiment Telescope)~\cite{Sreekumar:1997un}, and the Fermi Gamma-Ray Space Telescope, gradually lead to the conclusion that most of the observed gamma-ray emission could be attributed to known gamma-ray source classes (such as active galactic nuclei), although it remains possible that a non-negligible component of the high-latitude background could originate from dark matter~\cite{Ackermann:2015tah}. 

Motivated by their high densities of dark matter and low levels of baryonic activity, dwarf spheroidal galaxies -- satellites of the Milky Way -- have in recent years become a prime target of gamma-ray telescopes searching for evidence of dark matter annihilations. Fermi's study of dwarf galaxies has provided the strongest limits on the dark matter annihilation cross section to date, strongly constraining WIMPs lighter than $\sim$100 GeV or so in mass~\cite{Ackermann:2015zua}. Ground based gamma-ray telescopes have also used observations of dwarf galaxies to constrain the annihilations of heavier dark matter candidates. Although complicated by imperfectly understood backgrounds, gamma-ray observations of the Milky Way's Galactic Center are also highly sensitive to annihilating WIMPs. A significant excess of GeV-scale gamma-rays has been identified from this region, consistent with arising from the annihilations of $\sim$$50$ GeV particles~\cite{Goodenough:2009gk,Daylan:2014rsa}. An active debate is currently taking place regarding the interpretation of these observations. Alternative targets for indirect searches have also been proposed, including Galactic dark matter subhalos not associated with dwarf galaxies \cite{Diemand:2006ik,Pieri:2007ir}, and density ``spikes'' of dark matter around black holes \cite{Gondolo:1999ef,Zhao:2005zr,Bertone:2005xz}. 

%and the diffuse extragalactic gamma-ray background \cite{Ullio:2002pj}.

%It is in principle possible that such substructures have been detected as unassociated gamma-ray sources by the Fermi LAT satellite \cite{Bertoni:2015mla,Schoonenberg:2016aml}.    

Over approximately the same period of time, great progress has also been made in the measurement of the cosmic-ray antiproton spectrum, including successive advances by the CAPRICE~\cite{Boezio:1997ec,Boezio:2001ac}, BESS~\cite{Asaoka:2001fv,Abe:2008sh}, AMS~\cite{Aguilar:2002ad}, and PAMELA~\cite{Adriani:2010rc} experiments. When these measurements are combined with our current understanding of cosmic-ray production and propagation, they appear to indicate that the observed cosmic ray antiproton spectrum originates largely from conventional secondary production (cosmic-ray interactions with gas), although a significant contribution from dark matter remains a possibility. These measurements generally yield constraints on annihilating dark matter that are not much less stringent than those derived from gamma-ray observations. 

Compared to antiprotons, measurements of the cosmic-ray positron spectrum have been more difficult to interpret. Building upon earlier measurements~\cite{MullerTang,Golden:1995sq,Boezio:2000zz}, the balloon-bourne HEAT experiment observed in 1994, 1995, and 2000 indications of an excess of cosmic-ray positrons at energies above $\sim$10 GeV, relative to the rate predicted from standard secondary production~\cite{Barwick:1997ig}. This was later confirmed, and measured in much greater detail, by a series of space-based experiments: AMS~\cite{Aguilar:2007yf}, PAMELA~\cite{Adriani:2008zr}, and AMS-02~\cite{AMS02}. Although this positron excess received much attention as a possible signal of annihilating dark matter, this possibility is now strongly constrained by a variety of arguments (e.g.~Ref.~\cite{Bertone:2008xr,Galli:2009zc,Slatyer:2009yq}), and plausible astrophysical explanations have also been proposed (e.g.~Ref.~\cite{Hooper:2008kg}). 

As large volume neutrino telescopes began to be deployed, such experiments became increasingly sensitive to dark matter annihilating in the interiors of the Sun and Earth. The AMANDA detector at the South Pole~\cite{Ahrens:2002eb}, along with Super-Kamiokande in Japan~\cite{Desai:2004pq}, each significantly improved upon previous limits, to be followed most notably by IceCube~\cite{Aartsen:2012kia} and ANTARES~\cite{Adrian-Martinez:2013ayv}. Constraints from neutrino telescopes are currently competitive with those derived from direct detection experiments for the case of WIMPs with spin-dependent interactions with nuclei.

Many of the strategies employed to search for annihilating dark matter have also been used to constrain the rate at which dark matter particles might decay. In addition to constraints on gravitinos and other potentially unstable particles, such searches are particularly interesting within the context of sterile neutrino dark matter. Sterile neutrinos with masses in the range of $\sim$1-100 keV are predicted to decay (into an active neutrino and a photon) at a rate that could generate a potentially observable X-ray line~\cite{Abazajian:2001vt}. In fact, considering the standard case of Dodelson-Widrow production (as discussed in Chapter~\ref{particles}), the combination of constraints from X-ray observations and measurements of the Lyman-$\alpha$ forest~\cite{Viel:2005qj,Seljak:2006bg} disfavor sterile neutrino dark matter over this entire mass range. Models with enhanced production in the early Universe~\cite{Shi:1998km} can evade such constraints, however, and continue to receive considerable interest. In particular, reports of a 3.55 keV line observed from a collection of galaxy clusters~\cite{Bulbul:2014sua,Boyarsky:2014jta} have recently received a great deal of attention within the context of a decaying sterile neutrino.

\subsection{Axion Experiments}

For some time, there has been an active experimental program searching for dark matter axions, most notably in the form of the Axion Dark Matter eXperiment (ADMX). The idea behind this effort is to make use of the photon-photon-axion coupling, generically present in axion models, to convert dark matter axions in a strong and static magnetic field into a signal of nearly monochromatic microwave photons. This possibility was first suggested by Pierre Sikivie in 1983~\cite{Sikivie:1983ip}, and was later expanded upon by Sikivie~\cite{Sikivie:1985yu}, along with Lawrence Krauss, John Moody, Frank Wilczek and Donald Morris~\cite{Krauss:1985ub}. As the signal in such an experiment is maximized for a specific cavity frequency (corresponding to a specific axion mass), it is necessary that the resonant frequency of the cavity be tunable, making it possible to scan over a range of axion masses. 

The first laboratory constraints on dark matter axions were presented in the late 1980s, by a number of groups~\cite{DePanfilis:1987dk,Wuensch:1989sa,Hagmann:1990tj}. While the frequency range covered by these experiments was well suited to axion masses favored by dark matter abundance considerations (covering approximately $m_a \simeq 4.5-16.3 \, \mu$eV), their sensitivity was orders of magnitude below that required to test realistic axion models. In 2003, however, the ADMX Collaboration reported results that constrained realistic axion dark matter models, although only for a relatively narrow range of masses, $1.9-3.3 \, \mu$eV~\cite{Asztalos:2003px}. With anticipated upgrades~\cite{Asztalos:2009yp,Asztalos:2009yp}, ADMX is expected to be sensitive to a much larger range of axion masses and couplings, significantly constraining the axion dark matter parameter space in the coming years.

%\section*{Further reading} 
%\cite{1998ARA&A..36..599B}

%\include{indirect}
%\include{accelerators}
%\include{hints}

{\bf Acknowledgements.} We are grateful to the many pioneering physicists and astronomers who helped us to reconstruct the methods, ideas and circumstances that led to the establishment of the dark matter paradigm. In particular, we would to thank Andrea Biviano, Lars Bergstr{\"o}m, Albert Bosma, Shantanu Desai, Jaco de Swart, Jaan Einasto, John Ellis, Pierre Fayet, Ken Freeman, Katherine Freese, Steve Kent, Rocky Kolb, Keith Olive, Jerry Ostriker, Sergio Palomares-Ruiz, Jim Peebles, Joel Primack, Morton Roberts, Bernard Sadoulet, Bob Sanders, Gary Steigman, Alar Toomre, Juri Toomre, Scott Tremaine, Virginia Trimble, Michael Turner and Simon White. We also acknowledge many useful discussions with Jeroen van Dongen and Jaco de Swart. This work would not have been possible without the vast collection of articles and books, dating back to the 19th century, made freely available by NASA ADS and the Internet Archive Project. GB acknowledges support from the European Research Council through the ERC starting grant {\it WIMPs Kairos}. DH is supported by the US Department of Energy.

\appendix
\section{Further Reading} 
\begin{itemize}
\item Robert H.~Sanders. {\it The Dark Matter Problem: A Historical Perspective}. Cambridge University Press, 2010~\cite{2010dmp..book.....S}.
\item Virginia Trimble. {\it History of Dark Matter in Galaxies}, 2013~\cite{2013pss5.book.1091T}.
\item Jaan Einasto. {\it Dark matter and Cosmic Web Story}. World Scientific, 2014~\cite{2014dmcw.book.....E}.
\item H.~S.~Kragh. {\it Conceptions of Cosmos: From Myths to the Accelerating Universe}. Oxford University Press, 2006. 
\item  Andrea Biviano. {\it From Messier to Abell: 200 Years of Science with Galaxy Clusters}, 2000~\cite{2000cucg.confE...1B}.
\item Gianfranco Bertone, Dan Hooper and Joseph Silk. {\it Particle dark matter: Evidence, candidates and constraints}, 2004~\cite{Bertone:2004pz}. 
\item Lars Bergstr{\"o}m. {\it Nonbaryonic dark matter: Observational evidence and detection methods}, 2000~\cite{Bergstrom:2000pn}.
\item Gerard Jungman, Marc Kamionkowski and Kim Griest. {\it Supersymmetric Dark Matter}, 1996~\cite{Jungman:1995df}.
\item Edmund Bertschinger. {\it Simulations of Structure Formation in the Universe}, 1998~\cite{1998ARA&A..36..599B}.
\item Carlos Frenk and Simon White. {\it Dark matter and cosmic structure}, 2012 \cite{Frenk:2012ph}.
\item Jeremiah P.~Ostriker and Simon Mitton. {\it Heart of Darkness: Unraveling the Mysteries of the Invisible Universe}. Princeton University Press, 2013.

\end{itemize}

\bibliography{book2}

\end{document}